\documentclass[11pt,a4paper]{article}
\usepackage{amssymb}
\usepackage{amsmath}
\usepackage{framed}
\usepackage{slashed}
\usepackage{graphicx}
\usepackage{indentfirst}
\usepackage{a4wide}
\usepackage[utf8]{inputenc}
\usepackage{caption}
\usepackage{bbold}
\usepackage{braket}
\usepackage{cancel,pslatex}
\usepackage{color}
\usepackage[normalem]{ulem}
\usepackage{relsize}
\usepackage{adjustbox}
\usepackage{import}
\usepackage{bm}
\usepackage{array}
\usepackage{tikz}
\usetikzlibrary{math, shapes.geometric}
\usepackage{natbib}
\setcitestyle{authoryear,round}
\usepackage{epstopdf}
\usepackage{mathtools}
\usepackage{subcaption}
\usepackage{siunitx}
\usepackage{physics}
\usepackage{booktabs}

\usepackage{pgfplots}
\usepgfplotslibrary{groupplots,dateplot}
\usetikzlibrary{patterns,shapes.arrows}
\pgfplotsset{compat=newest}

\renewcommand{\vec}[1]{\bm{#1}}

\DeclareMathOperator{\diag}{diag}

\pgfplotsset{every axis/.append style={
    label style={font=\tiny},
    tick label style={font=\tiny},
    legend style={font=\tiny},
    scaled y ticks=false,
  }, yticklabel style={
    log ticks with fixed point
  }}
\pgfplotsset{}

\begin{document}
\title{\bf Determining the optimal focusing parameter in sparse promoting inversions of EMI surveys}
\author{{Wouter Deleersnyder~$^{a,b}$\thanks{wouter.deleersnyder@kuleuven.be}}, {David Dudal~$^{a}$\thanks{david.dudal@kuleuven.be (corresponding author)}, Benjamin Maveau~$^{a}$\thanks{benjamin.maveau@gmail.com}, Marieke Paepen~$^{b}$\thanks{marieke.paepen@ugent.be}}\\
\textit{{\small $^a$ KU Leuven Campus Kortrijk -- Kulak, Department of Physics, Etienne Sabbelaan 53 bus 7657,}}\\
\textit{{\small 8500 Kortrijk, Belgium}}\\
\textit{{\small $^b$           Ghent University, Department of Geology, Krijgslaan 281-S8, 9000 Gent, Belgium}}}

\date{}
\maketitle

\begin{abstract}
If the magnetic field caused by a magnetic dipole is measured, the electrical conductivity of the subsurface can be determined by solving the inverse problem. For this problem a form of regularisation is required as the {inverse problem} is ill-posed.  Commonly, Tikhonov regularisation is used which adds the $\ell_2$-norm of {(some difference between)} the model parameters to the objective function.
  As a result, a smooth conductivity profile is {promoted} and these types of inversions are very stable.
  However, it fails to recover sharp conductivity boundaries when the true profile has discontinuities, {as} the discontinuities cause oscillations in the obtained model parameters.
  To circumvent this problem, $\ell_0$-approximating norms can be used to promote discontinuous model parameters.
  Two of these {measures} are considered, the Minimum Gradient Support and the Cauchy {measure}.
  However, both {measures} contain a parameter which {interpolates} the function from the $\ell_2$- to the $\ell_0$-norm {or  $\ell_1$-norm}.
  To find the optimal value of this parameter, a new method{, not relying on any prior input,} is suggested.
  It is based on the $L$-curve method and finds a good balance between a continuous and discontinuous profile.
  The method is tested on synthetic data and is able to produce a conductivity profile similar to the true profile.
  Furthermore, the strategy is applied to newly acquired field data and the obtained profiles are in agreement with the results of other surveys at the same location.
  Finally, despite the fact that the Cauchy {measure} is only occasionally used to the best of our knowledge, we find that it performs at least as well as the Minimum Gradient Support measure, especially in combination with convex optimisation algorithms.
\end{abstract}


\section{Introduction}
Frequency domain electromagnetic induction (FDEM) uses a coil with an alternating current as a source while, at a distance, a second coil measures the electromagnetic response of the subsurface conductive target.
This magnetic field allows one to estimate the electrical conductivity profile of the soil, albeit that the reconstructed conductivity is not unique in general.
The profile can be used to monitor chemical pollution \citep{martinelli_laterally_2008,deidda_frequency-domain_2022}, search for {subsurface} structures {in archaeological} \citep{saey_electrical_2012} {and landfill mapping} \citep{deleersnyder2023estimating}, or detect saltwater intrusion \citep{scudiero_constrained_2011,rulff2026computational}.
All of this is done in a non-destructive way and, because it is a non-contacting method, measurements are done rather easily.

To transform the measurements (i.e.\ the magnetic field) to a conductivity profile, we need to solve the so called inverse problem which, in essence, is an optimisation problem.
The objective function that needs to be minimised contains {the data misfit between the observed and predicted data} from the model parameters, the conductivity profile in our case.
This optimisation is highly unstable, creating solutions which are strongly dependent on the measurement error.
Moreover, the solution is hardly unique.
We therefore use a process called regularisation, which alters our objective function to prefer geophysical probable solutions.
Tikhonov regularisation \citep{tikhonov_stability_1943} is frequently used, leading to a smooth conductivity profile.

However, there are multiple reasons why a piecewise constant profile is more appropriate.
The number of data points in electromagnetic induction (EMI) surveys is small, reducing the resolution and making it less sensitive to a rapid change in conductivity.
A piecewise constant profile allows a large change in conductivity.
Combining this with a fine resolution in the inverted profile, we can give a good estimate on the location of the interface.
For smooth profiles on the contrary, the profile will show a transition zone between layers and the interface depth will be subject to interpretation.
Different methods exist to abandon smooth model parameters \citep{farquharson_constructing_2007,paasche_cooperative_2007,hermans_imaging_2012,deleersnyder_inversion_2021,deleersnyder2023flexible,thibaut_new_2021} but they either require prior knowledge or fixing (a) parameter(s) in the inversion.
As the $\ell_{2}$-norm causes oscillations at discontinuities \citep{farquharson_non-linear_1998,loke_comparison_2003}, most of the previous mentioned methods abandon this norm in favour of the $\ell_{0}$- and $\ell_{1}$-norm\footnote{Most often, the $\ell_{0}$-norm of a vector $\vec x = (x_1,\ldots,x_n)$ is seen as the number of non-zero elements $x_i$. This definition, however, does not obey all the mathematical requirements of a proper norm. For completeness, we recall that in general, for $p\geq 1$, the $\ell_p$-norm of $\vec x$ is defined via $||\vec x||_p=\left(|x_1|^p+\ldots |x_n|^p\right)^{1/p}$.}.
These are much more suited for the desired model parameters {in some contexts}, but these norms are not {differentiable}.
Several, {differentiable}, approximations to these norms exist such as the Minimum Support~\citep{last_compact_1983} or the Cauchy log {measure} \citep{guitton_blocky_2012}, see Equations~\ref{normMGS}-\ref{normCauchy}.

{All these measures contain a focusing parameter which balances between smooth and piecewise constant/smooth functions. The popularity of these measures has increased
in recent years \citep{deidda_inversion_2020, deidda2018smooth, klose_laterally_2022}. The sheer importance of the focusing parameter was already established in~\citep{blaschek_new_2008}: a too small value causes small changes in conductivity to be ignored while a too large value will cause oscillations at discontinuities. Despite these known issues, the parameter is often set a priori, and few methods exist which determine it automatically. Earlier suggested methods are via the point of
maximum curvature \citep{zhdanov2004minimum}, interpolation based on
model parameters \citep{kim_time-lapse_2011} or even the mean value of the model parameters \citep{carbajal2012focused}. We refer to e.g.~\citep{fiandaca_generalized_2015} for a short overview of methods to determine the focusing parameter.}

{Other methods exist to abandon smooth model
parameters \citep{deleersnyder_inversion_2021,farquharson_constructing_2007,hermans_imaging_2012,paasche_cooperative_2007,thibaut_new_2021} but they either require prior
knowledge or fixing parameter(s) in the inversion. }

In this paper, we therefore suggest a new method to {objectively} determine the focusing parameter, based on a scheme similar to the $L$-curve method \citep{hansen1999curve}{, which yields an estimation of the parameter that avoids the aforementioned unwanted effects in the model parameters. Moreover, no prior information is necessary for the proposed scheme to work.} {We will not only consider the commonly used Minimum Gradient Support {measures}, but also make a case for the Cauchy {measure} in this context. Indeed, the latter {measure} shows quite some flexibility in producing blockyness
whilst still allowing for some smoothness, an interesting asset if the available a priori knowledge of the subsurface is incomplete.}

We first develop our new inversion methodology for the 1D low-frequency EM setting. In the Conclusion, we will briefly touch upon a few scenario's to extend our findings to the 2D or 3D case, for which robust inversion strategies are way more challenging.

We then apply our method on four synthetic conductivity profiles.
The first one is a simple three layer soil profile and we use a smoothened version as the second profile.
The third synthetic profile is generated from the conductivity profile obtained from borehole logging at Hermalle-sous-Argenteau (Belgium) along the river Meuse. The final synthetic case considers a resistive layer embedded in a conductive background, and mimics a survey with the EM34 (Geonics) instrument.
Finally we apply it on data collected at De Westhoek, a nature reserve in De Panne (Belgium).
At this site, saltwater intrusion causes a sharp conductivity spike in the profile, which we are effectively able to locate with our new method.

\section{Inversion method}
\label{sec:inversion-method}
Our inverse problem consists of minimising the objective function
\begin{equation}
  \label{eq:73}
  \phi(\vb{m}) = \left\lVert \frac{1}{N}\vb{W}_{d}\vdot(\vb{F}(\vb{m}) - \vb{d})\right\rVert^2_2 + \lambda \phi_{m}(\vb{W}_{m}\vdot \log(\vb{m})),
\end{equation}
where $\vb{W}_{d,m}$ are weighing matrices, $\vb{F}$ is the forward model, $\vb{d}$ is the{ observed} data of size $N$ and $\phi_{m}$ is the regularisation function.
The latter cures the ill-posedness of the optimisation problem which is caused by the measurement error $\vb{\eta}$ and the non-linearity of the forward model.
We set $\vb{W}_{d}$ equal to $\diag{(\vb*{\eta}^{-1})}$ and choose the regularisation parameter $\lambda$ so that
\[\phi_{d}=\left\lVert \frac{1}{N}\vb{W}_{d}\vdot(\vb{F}(\vb{m}) - \vb{d})\right\rVert^2_2\] is slightly larger than 1.
This is based on the discrepancy principle \citep{morozov_solution_1966}, stating that the mismatch between $\vb{F}$ and $\vb{d}$ must be equal or larger than the error $\vb*{\eta}$ on the data.
Finally, the matrix $\vb{W}_{m}$ can be any matrix, but is commonly the identity matrix or a matrix implementing a discretization of the derivative.

\subsection{The forward model}
  \label{subsec:the_forward_model}
As forward model in Equation~\eqref{eq:73} the one dimensional damped model is used \citep{delrue_damped_2020}.
  For the underlying mathematical details we refer to \citep{delrue_damped_2020}.
  This model is an approximation to the exact result of \citep{keller_electrical_1966}\footnote{For the record, we recall here that the exact model is one-dimensional, assuming a layered subsurface with axial symmetry. This 1D model is not known in closed form expression, but it can be dealt with numerically.} and is quasi-linear.
That is, the forward problem can be written as a matrix equation, but the matrix itself depends on the conductivity profile.
With respect to the low induction number model~\citep{wait_note_1962}, it has the advantage to be valid for larger induction numbers $\mu\omega\sigma s^{2}$, where $\mu$ is the magnetic permeability, $\omega$ the angular frequency of the electromagnetic wave, $\sigma$ the conductivity of the soil and $s$ the distance between the emitter and receiver coils of the signal.
This forward model is very fast to compute with and its validity has been extensively tested in \citep{delrue_damped_2020}.
As we will construct an L-curve, a fast forward model is an asset for the numerous forward runs needed.
Its validity in highly conductive circumstances is also crucial since we test our method on data in a salt water intruded area. Any inversion based on a forward model that assumes a low induction number (LIN) will definitely render untrustworthy results.

The generation of this signal can be due to two orientations, either a horizontal or vertical current loop causing a vertical or horizontal dipole respectively.
  For the generation of our synthetic data with the exact result of \citep{keller_electrical_1966}, we assume a horizontal current loop  and measure with a horizontal current loop (Horizontal Coplanar, HCP) and a vertical loop (Perpendicular PRP).
  The actual data measured at nature reserve De Westhoek contains both the HCP configuration and the Vertical Coplanar (VCP) orientation.
  In the latter case, both the emitter and receiver coils are vertical and parallel.

\subsection{Solving the optimisation problem}
\label{subsec:solving_the_optmization}
We solve this optimisation problem using the Gauss-Newton algorithm, which requires the derivative
\begin{equation}
  \label{eq:91}
  \grad_{\vb{m}}{\phi(\vb{m})} = \vb{J}^{\dagger}\vdot\vb*{\Psi}\vdot\vb{r},
\end{equation}
where we defined the following symbols:
\begin{equation}
  \label{eq:3}
  \vb{J} = \mqty( \vb{W}_{d}\vb{G}_{\vb{F}} \\ \sqrt{\lambda}\vb{W}_{m}),\qquad
  \vb{r} = \mqty(\vb{W}_{d}\vb{r}_{d} \\ \sqrt{\lambda}\vb{W}_{m}\vb{m}),\qquad
  \vb*{\Psi} = \mqty(\text{diag}\left(\frac{\phi_{d}^{\prime}(\vb{W}_{d}\vdot\vb{r}_{d})}{\vb{W}_{d}\vdot\vb{r}_{d}}\right) & \vb{0} \\ \vb{0} & \text{diag}\left(\frac{\phi_{m}^{\prime}(\vb{W}_{m}\vdot\vb{m})}{\vb{W}_{m}\vdot\vb{m}}\right)),
\end{equation}
$\vb{G}_{\vb{F}}$ is the gradient matrix of the forward function and $\phi_{d,m}^{\prime}$ is the derivative of the respective {measure} function with respect to its argument.
Taking the derivative of the gradient results in the approximate Hessian $\vb{H}[\phi(\vb{m})] = \vb{J}^{\dagger}\vdot\vb*{\Psi}\vdot \vb{J}$, where we neglected the second order derivatives.
These have a small contribution as long as the residual $\vb{r}$ is small \citep{nocedal_numerical_2006}, which should be the case if we are close enough to a minimum.
To find a direction where the objective function is smaller, we evaluate it at $\vb{m}+\delta\vb{m}$, and use a Taylor expansion around $\vb{m}$ {wherein the aforementioned approximate Hessian is substituted, to write}
\begin{equation}
  \phi(\vb{m} + \delta\vb{m}) = \phi(\vb{m}) + \grad_{\vb{m}}{\phi(\vb{m})}\vdot\delta\vb{m} + \delta \vb{m}\vdot \vb{H}[\phi(\vb{m})]\vdot\delta \vb{m}.
\end{equation}

The minimum of this function can readily be found if we neglect the third order in $\delta \vb{m}$.
Indeed, we reduce our objective function to a quadratic equation in $\delta \vb{m}$, whose minimum is equal to the solution of the following equation
\begin{equation}
  \label{eq:2}
  (\vb{J}^\dagger\vdot \vb*{\Psi} \vdot \vb{J}) \cdot\delta\vb{m} = -\grad{\phi(\vb{m})}.
\end{equation}

To increase performance, we modify the algorithm in two ways.
First, we demand a positive electrical conductivity, further reducing the model space.
To accomplish this, we use the Gradient projection reduced Newton algorithm~\citep{vogel_computational_2002}\footnote{By working in log space, one can also force the inversion towards a positive conductivity. We, however, noticed no improvement in speed or accuracy and we therefore used a bounded solver.}.
Secondly, we also perform a line-search with the step determined from equation~\eqref{eq:2}.
Using the algorithm from \citep{more_line_1994}, we can assure that the first Wolfe condition is obeyed \citep{wolfe_convergence_1969}.

As already said, Tikhonov {regularisation} \citep{tikhonov_stability_1943} is probably the most common and simplest form of {regularisation}.
It uses the $\ell_{2}$-norm as $\phi_{m}$, resulting in smooth model parameters.
Due to the nature of our problem, we, however, rather expect piecewise constant/smooth model parameters.
Mathematically, this translates in model parameters for which the differences between subsequent values are small apart from a few at the interfaces.
Using the matrix
\begin{equation*}
  \vb{W}_{m} = \mqty(\dmat{1 & -1 & 0 & \\0 & 1 & -1, \cdots, 1 & -1 & 0\\ 0 & 1 & -1})
\end{equation*}
as weighing matrix, {we get in Eq.~\eqref{eq:73} a vector $\vb{W}_{m}\vdot\vb{m}$ containing the differences between adjacent layers conductivities. The sparser this vector is, the more piecewise constant like the conductivity profile $\vb{m}$ will be}. In this case an $\ell_{0}$-norm for the regularizing $\phi_m$ is preferred, {as e.g.~an $\ell_2$-norm would try to wash out again the jumps between the piecewise constant parts.}
\begin{figure}[t]
  \centering
  \includegraphics[width=\linewidth]{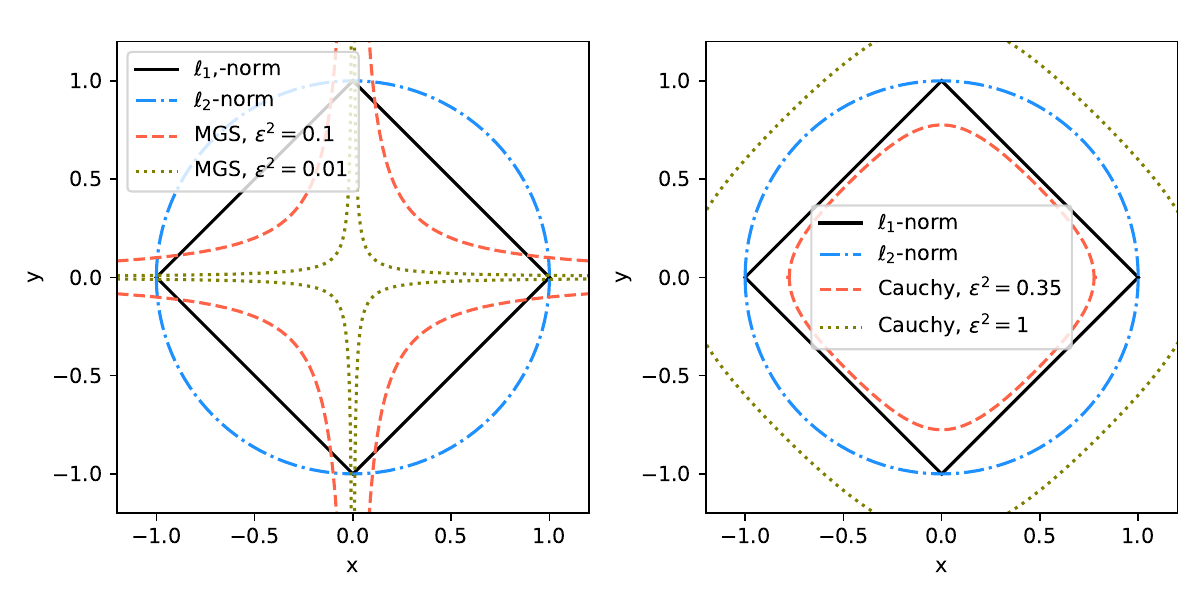}
  \caption{The unit circle/diamond/astroid for the different norms or measures in two dimensions. For smaller $\epsilon$, the measures tend to more effectively promote sparsity.}
  \label{fig:norms}
\end{figure}

Unfortunately, the derivative of the $\ell_0$-norm does not exist as the function is discontinuous, while the cusps in the $\ell_1$-norm also prohibits proper differentiation.
This is circumvented by using a smooth function which approximates the $\ell_{0}$-norm.
While different approximating functions (e.g.~Cauchy or Minimum (Gradient) Support---M(G)S\footnote{Depending on the input of this function, it is either called the Minimum Gradient Support or the Minimum Support. The former takes the differences between conductivity layers as input (as in our case), while the latter uses the conductivity of the layers. Needless to say, many other penalty functions exist and could be tested, see for example \citep{feng2020modified}, but this is not the goal of the current paper.}) exist, all depend on a focusing parameter $\epsilon$.
This focusing parameter balances between smooth and blocky model parameters and its importance cannot be overestimated.
We used
\begin{eqnarray}
  \text{MGS: } && \phi_m(\vec{x}) = \frac{\vec{x}^2}{\vec{x}^2 + \epsilon^2}\qquad\text{and}\label{normMGS}\\
  \text{Cauchy: } && \phi_m(\vec{x}) = \ln(1 + \frac{\vec{x}^2}{\epsilon^2}), \label{normCauchy}
\end{eqnarray}
as both can get very blocky and smooth model parameters for small and large $\epsilon$ respectively. In Figure~\ref{fig:norms}, these two norm {approximations} are illustrated  together with the $\ell_{1}$- and $\ell_{2}$-norm for a two-dimensio\-nal vector $\vec{x}=(x,y)$ lying on the respective unit ``circles''.
For small values of $|\vec{x}|$ with respect to the focusing parameter, the two {forms} {both converge quite fast to the} $\ell_{2}$-norm. They therefore are both able to produce smooth results. For smaller focusing parameters, they are both less-penalising than the $\ell_{1}$-norm.  The Cauchy function keeps increasing for larger changes in the $\vec x$ components , albeit at a slower rate.
The MGS, however, converges to the value of 1. Hence both {measures} penalize multiple changes in favor of one big change. The MGS {measure} is stricter from this viewpoint as it penalizes all large changes with the same value (namely 1), while the Cauchy {measure} still prefers one major change in favor of multiple ones. We can also use the $\ell_1$-norm as a baseline, as this norm is the same for multiple small changes or one big change, it is just the sum of the absolute values of the changes. For the MGS case, in particular for small $\epsilon$, the measure is $\approx 1$, as it is clear from  \eqref{normMGS}. Moreover, the variations around that 1 are small for small $\vec x^2$, the latter most easily realizable with a vector $\vec x$ with as least as possible components.

On the other hand, the Cauchy expression \eqref{normCauchy} for small $\epsilon$ will prefer a small $\vec x^2$ too, leading to a similar conclusion as for the MGS norm.  That is, both prefer sparsity.

We will use the discrepancy principle to set the regularisation parameter $\lambda$ but still need to determine the optimal focusing parameter $\epsilon$.
We will discuss a novel approach to find the optimal value of the focusing parameter {objectively}.

\subsection{Selecting the optimal focusing parameter}
\label{subsec:selecting_the_optimal_focusing}

\begin{figure}
  \centering
  \begin{tikzpicture}[node distance=2cm,every text node part/.style={align=center}]
    \tikzstyle{startstop} = [rectangle, rounded corners, minimum width=3cm, minimum height=1cm,text centered, draw=black, fill=red!30]
    \tikzstyle{io} = [trapezium, trapezium left angle=70, trapezium right angle=110, minimum width=3cm, minimum height=1cm, text centered, draw=black, fill=blue!30]
    \tikzstyle{process} = [rectangle, minimum width=3cm, minimum height=1cm, text centered, draw=black, fill=orange!30]
    \tikzstyle{decision} = [diamond, minimum width=3cm, minimum height=1cm, text centered, draw=black, fill=green!30]
    \tikzstyle{arrow} = [thick,->,>=stealth]

    \node (start) [startstop] {Random initial\\vector $\vb{m}_{0}$};
    \node (it0) [process, below of=start] {$\min\limits_{\vb{m}} \phi(\vb{m}_{i} = \vb{m}_{0}, \epsilon_{0}) = \vb{m}_{\epsilon_{0}}$};
    \node (st_or_re) [decision, below of=it0] {Strategy};

    \node (st) [process, below of=st_or_re, yshift=-.5cm] {$\min\limits_{\vb{m}} \phi(\vb{m}_{i} = \vb{m}_{\epsilon_{0}}, \epsilon_{1}) = \vb{m}_{\epsilon_{1}}$};
    \node (st2) [below of=st, yshift=.5cm] {\vdots};
    \node (st3) [process, yshift=.5cm, below of=st2] {$\min\limits_{\vb{m}} \phi(\vb{m}_{i} = \vb{m}_{\epsilon_{0}}, \epsilon_{N}) = \vb{m}_{\epsilon_{N}}$};

    \node (re) [process, right of=st_or_re, xshift=4cm] {$\min\limits_{\vb{m}} \phi(\vb{m}_{i} = \vb{m}_{\epsilon_{0}}, \epsilon_{1}) = \vb{m}_{\epsilon_{1}}$};
    \node (re2) [process, below of=re, yshift=-.5cm] {$\min\limits_{\vb{m}} \phi(\vb{m}_{i} = \vb{m}_{\epsilon_{1}}, \epsilon_{2}) = \vb{m}_{\epsilon_{2}}$};
    \node (re3) [below of=re2, yshift=.5cm] {\vdots};
    \node (re4) [process, below of=re3, yshift=.5cm,] {$\min\limits_{\vb{m}} \phi(\vb{m}_{i} = \vb{m}_{\epsilon_{N-1}}, \epsilon_{N}) = \vb{m}_{\epsilon_{N}}$};

    \draw[arrow] (start) -- (it0);
    \draw[arrow] (it0) -- (st_or_re);
    \draw[arrow] (st_or_re) -- (st) node [midway, left] {start};
    \draw[arrow] (st) -- (st2);
    \draw[arrow] (st2) -- (st3);
    \draw[arrow] (st_or_re) -- (re) node [midway, above] {reuse};
    \draw[arrow] (re) -- (re2);
    \draw[arrow] (re2) -- (re3);
    \draw[arrow] (re3) -- (re4);
  \end{tikzpicture}
  \caption{A flowchart to illustrate the start and reuse strategy. The argument $\vb{m}_{i}$ of $\phi_{m}$ is the starting model parameter of the inversion. While the left branch (the start strategy) is shown sequentially, the subsequent inversions can be solved in parallel since they are independent on the previous solution.}
  \label{fig:diagram_start_reuse}
\end{figure}

Using our optimisation algorithm described above, we solve the inverse problem for different values of the focusing parameter. These values are uniformly chosen in log-space between $10^{0}$ and $10^{-5}$.

\begin{itemize}
\item We first solve the problem for the largest value of the focusing parameter, then we use this solution as starting point for all the other solvers {(and thus the regularisation parameter remains fixed)}.
This prevents the problem of multi-modality, which arises for smaller values of the focusing parameter. {Moreover, this will reduce the number of required iterations for each individual optimisation problem, and speeds up the procedure.}
We call this the \emph{start strategy}.
As the method only requires the solution of the inversion with a large focusing parameter, we can easily compute the other results in parallel.
\item Another approach uses the solution of the previous focusing parameter to get the next result.
This method is called the \emph{reuse strategy}.
This causes a smoother and more stable $L$-curve, but it tends to flatten peaks in the conductivity profile.
The latter is illustrated in the blocky 3-layer soil profile of Subsection~\ref{sec:synth_data}.
\end{itemize}

\begin{figure}[ht]
  \centering
  \subcaptionbox{}{
\begin{tikzpicture}

\definecolor{darkgray176}{RGB}{176,176,176}
\definecolor{darkviolet1910191}{RGB}{191,0,191}
\definecolor{goldenrod1911910}{RGB}{191,191,0}
\definecolor{lightgray204}{RGB}{204,204,204}

\begin{axis}[
height=.3\textwidth,
legend cell align={left},
legend style={fill opacity=0.8, draw opacity=1, text opacity=1, draw=lightgray204},
log basis y={10},
tick align=outside,
tick pos=left,
width=.99\textwidth,
x grid style={darkgray176},
xlabel={d (m)},
xmin=-0.55, xmax=11.55,
xtick style={color=black},
y grid style={darkgray176},
ylabel={\(\displaystyle \sigma\) (mSm\(\displaystyle ^{-1}\))},
ymin=0.05, ymax=1,
ymode=log,
ytick style={color=black},
ytick={0.05,0.1,0.5,1},
yticklabels={50,100,500,1000}
]
\addplot [semithick, black]
table {%
0 0.1
2.5 0.1
2.5 0.8
5 0.8
5 0.1
10 0.1
10 0.1
11 0.1
};
\addlegendentry{True}
\addplot [semithick, darkviolet1910191, dotted]
table {%
0 0.0868874567372207
0.526315789473684 0.0868874567372207
0.526315789473684 0.0832165948227497
1.05263157894737 0.0832165948227497
1.05263157894737 0.0849487472322219
1.57894736842105 0.0849487472322219
1.57894736842105 0.0897053351433892
2.10526315789474 0.0897053351433892
2.10526315789474 0.700819643822206
2.63157894736842 0.700819643822206
2.63157894736842 0.721235640534541
3.15789473684211 0.721235640534541
3.15789473684211 0.696080901603888
3.68421052631579 0.696080901603888
3.68421052631579 0.625832905700782
4.21052631578947 0.625832905700782
4.21052631578947 0.523199682148318
4.73684210526316 0.523199682148318
4.73684210526316 0.407308913722398
5.26315789473684 0.407308913722398
5.26315789473684 0.301427824618332
5.78947368421053 0.301427824618332
5.78947368421053 0.226693303411575
6.31578947368421 0.226693303411575
6.31578947368421 0.176956913205394
6.84210526315789 0.176956913205394
6.84210526315789 0.142569148587891
7.36842105263158 0.142569148587891
7.36842105263158 0.11782049345836
7.89473684210526 0.11782049345836
7.89473684210526 0.0993960647876424
8.42105263157895 0.0993960647876424
8.42105263157895 0.0852879210679224
8.94736842105263 0.0852879210679224
8.94736842105263 0.0742254841554418
9.47368421052632 0.0742254841554418
9.47368421052632 0.0653736153444614
10 0.0653736153444614
10 0.0581648370204388
11 0.0581648370204388
};
\addlegendentry{$\epsilon=1.6e-01>\epsilon_c$}
\addplot [semithick, blue, dashed]
table {%
0 0.087609117906306
0.526315789473684 0.087609117906306
0.526315789473684 0.0847620478421255
1.05263157894737 0.0847620478421255
1.05263157894737 0.0859320502344069
1.57894736842105 0.0859320502344069
1.57894736842105 0.0895671508450695
2.10526315789474 0.0895671508450695
2.10526315789474 0.629412077567342
2.63157894736842 0.629412077567342
2.63157894736842 0.672045061646295
3.15789473684211 0.672045061646295
3.15789473684211 0.706254779950321
3.68421052631579 0.706254779950321
3.68421052631579 0.722402738529109
4.21052631578947 0.722402738529109
4.21052631578947 0.718152266567957
4.73684210526316 0.718152266567957
4.73684210526316 0.696002237208224
5.26315789473684 0.696002237208224
5.26315789473684 0.0955898643858094
5.78947368421053 0.0955898643858094
5.78947368421053 0.0906397397185429
6.31578947368421 0.0906397397185429
6.31578947368421 0.0858696662034068
6.84210526315789 0.0858696662034068
6.84210526315789 0.081354960012311
7.36842105263158 0.081354960012311
7.36842105263158 0.0771407680137099
7.89473684210526 0.0771407680137099
7.89473684210526 0.0732478263505798
8.42105263157895 0.0732478263505798
8.42105263157895 0.0696788656514016
8.94736842105263 0.0696788656514016
8.94736842105263 0.0664244101427341
9.47368421052632 0.0664244101427341
9.47368421052632 0.0634674458937306
10 0.0634674458937306
10 0.0607868746106451
11 0.0607868746106451
};
\addlegendentry{$\epsilon=1.4e-01=\epsilon_c$, start}
\addplot [semithick, goldenrod1911910, dashed]
table {%
0 0.0878741527397826
0.526315789473684 0.0878741527397826
0.526315789473684 0.0869764979990246
1.05263157894737 0.0869764979990246
1.05263157894737 0.0884027874781238
1.57894736842105 0.0884027874781238
1.57894736842105 0.0910581726012711
2.10526315789474 0.0910581726012711
2.10526315789474 0.628336811345192
2.63157894736842 0.628336811345192
2.63157894736842 0.647128639694423
3.15789473684211 0.647128639694423
3.15789473684211 0.65812430714756
3.68421052631579 0.65812430714756
3.68421052631579 0.660001035825451
4.21052631578947 0.660001035825451
4.21052631578947 0.653794185982952
4.73684210526316 0.653794185982952
4.73684210526316 0.641671856423061
5.26315789473684 0.641671856423061
5.26315789473684 0.626195499897184
5.78947368421053 0.626195499897184
5.78947368421053 0.0562160713868449
6.31578947368421 0.0562160713868449
6.31578947368421 0.0547507650704529
6.84210526315789 0.0547507650704529
6.84210526315789 0.0533422008902618
7.36842105263158 0.0533422008902618
7.36842105263158 0.0519996793713607
7.89473684210526 0.0519996793713607
7.89473684210526 0.0507289885372542
8.42105263157895 0.0507289885372542
8.42105263157895 0.0495331183588796
8.94736842105263 0.0495331183588796
8.94736842105263 0.048412863653932
9.47368421052632 0.048412863653932
9.47368421052632 0.0473673685636454
10 0.0473673685636454
10 0.046394619171991
11 0.046394619171991
};
\addlegendentry{$\epsilon=1.1e-01=\epsilon_c$, reuse}
\addplot [semithick, red, dashed]
table {%
0 0.0731678454464712
0.526315789473684 0.0731678454464712
0.526315789473684 0.0731678453869087
1.05263157894737 0.0731678453869087
1.05263157894737 0.0731678453735069
1.57894736842105 0.0731678453735069
1.57894736842105 0.519888884891878
2.10526315789474 0.519888884891878
2.10526315789474 0.519888885772944
2.63157894736842 0.519888885772944
2.63157894736842 0.519888886858427
3.15789473684211 0.519888886858427
3.15789473684211 0.519888887877059
3.68421052631579 0.519888887877059
3.68421052631579 0.519888888673802
4.21052631578947 0.519888888673802
4.21052631578947 0.519888889204088
4.73684210526316 0.519888889204088
4.73684210526316 0.51988888947183
5.26315789473684 0.51988888947183
5.26315789473684 0.519888889503549
5.78947368421053 0.519888889503549
5.78947368421053 0.091895533659907
6.31578947368421 0.091895533659907
6.31578947368421 0.0918955336246489
6.84210526315789 0.0918955336246489
6.84210526315789 0.0918955335851088
7.36842105263158 0.0918955335851088
7.36842105263158 0.0918955335413167
7.89473684210526 0.0918955335413167
7.89473684210526 0.0918955334990468
8.42105263157895 0.0918955334990468
8.42105263157895 0.0918955334571838
8.94736842105263 0.0918955334571838
8.94736842105263 0.0918955334179246
9.47368421052632 0.0918955334179246
9.47368421052632 0.0918955333793905
10 0.0918955333793905
10 0.0918955333424722
11 0.0918955333424722
};
\addlegendentry{$\epsilon=1.0e-05\ll\epsilon_c$, reuse}
\end{axis}

\end{tikzpicture}}
  \subcaptionbox{}{
\begin{tikzpicture}

\definecolor{darkgray176}{RGB}{176,176,176}
\definecolor{darkviolet1910191}{RGB}{191,0,191}
\definecolor{goldenrod1911910}{RGB}{191,191,0}
\definecolor{lightgray204}{RGB}{204,204,204}

\begin{axis}[
height=.3\textwidth,
legend cell align={left},
legend style={fill opacity=0.8, draw opacity=1, text opacity=1, draw=lightgray204},
log basis y={10},
tick align=outside,
tick pos=left,
width=.99\textwidth,
x grid style={darkgray176},
xlabel={d (m)},
xmin=-0.55, xmax=11.55,
xtick style={color=black},
y grid style={darkgray176},
ylabel={\(\displaystyle \sigma (\si{\milli\siemens\per\meter})\)},
ymin=0.05, ymax=1,
ymode=log,
ytick style={color=black},
ytick={0.05,0.1,0.5,1},
yticklabels={50,100,500,1000}
]
\addplot [semithick, black]
table {%
0 0.1
2.5 0.1
2.5 0.8
5 0.8
5 0.1
10 0.1
10 0.1
11 0.1
};
\addlegendentry{True}
\addplot [semithick, darkviolet1910191, dotted]
table {%
0 0.0871874014782372
0.526315789473684 0.0871874014782372
0.526315789473684 0.0877799389332087
1.05263157894737 0.0877799389332087
1.05263157894737 0.0890082400734779
1.57894736842105 0.0890082400734779
1.57894736842105 0.0899997002157021
2.10526315789474 0.0899997002157021
2.10526315789474 0.76366222873453
2.63157894736842 0.76366222873453
2.63157894736842 0.719317716077048
3.15789473684211 0.719317716077048
3.15789473684211 0.637902560319236
3.68421052631579 0.637902560319236
3.68421052631579 0.534634658156205
4.21052631578947 0.534634658156205
4.21052631578947 0.425526431481449
4.73684210526316 0.425526431481449
4.73684210526316 0.33462049910983
5.26315789473684 0.33462049910983
5.26315789473684 0.272237788122279
5.78947368421053 0.272237788122279
5.78947368421053 0.227706948899299
6.31578947368421 0.227706948899299
6.31578947368421 0.194763425319821
6.84210526315789 0.194763425319821
6.84210526315789 0.16967972189813
7.36842105263158 0.16967972189813
7.36842105263158 0.150123948076797
7.89473684210526 0.150123948076797
7.89473684210526 0.134573707933512
8.42105263157895 0.134573707933512
8.42105263157895 0.121999244219935
8.94736842105263 0.121999244219935
8.94736842105263 0.111682830560175
9.47368421052632 0.111682830560175
9.47368421052632 0.10311164964233
10 0.10311164964233
10 0.0959106729920237
11 0.0959106729920237
};
\addlegendentry{$\epsilon=2.0e-01>\epsilon_c$}
\addplot [semithick, blue, dashed]
table {%
0 0.072671149716147
0.526315789473684 0.072671149716147
0.526315789473684 0.0715742701826133
1.05263157894737 0.0715742701826133
1.05263157894737 0.0711388847631433
1.57894736842105 0.0711388847631433
1.57894736842105 0.510554166609597
2.10526315789474 0.510554166609597
2.10526315789474 0.523421928265589
2.63157894736842 0.523421928265589
2.63157894736842 0.540961204653078
3.15789473684211 0.540961204653078
3.15789473684211 0.558126724767106
3.68421052631579 0.558126724767106
3.68421052631579 0.571941736037987
4.21052631578947 0.571941736037987
4.21052631578947 0.581010569878072
4.73684210526316 0.581010569878072
4.73684210526316 0.585004697126048
5.26315789473684 0.585004697126048
5.26315789473684 0.10824063783668
5.78947368421053 0.10824063783668
5.78947368421053 0.107962007642161
6.31578947368421 0.107962007642161
6.31578947368421 0.10757159288735
6.84210526315789 0.10757159288735
6.84210526315789 0.107097849922875
7.36842105263158 0.107097849922875
7.36842105263158 0.106567031286516
7.89473684210526 0.106567031286516
7.89473684210526 0.106002113617231
8.42105263157895 0.106002113617231
8.42105263157895 0.10542242312848
8.94736842105263 0.10542242312848
8.94736842105263 0.10484369256431
9.47368421052632 0.10484369256431
9.47368421052632 0.104278354674583
10 0.104278354674583
10 0.103735933932057
11 0.103735933932057
};
\addlegendentry{$\epsilon=1.2e-01\epsilon_c$, start}
\addplot [semithick, goldenrod1911910, dashed]
table {%
0 0.0837430524118706
0.526315789473684 0.0837430524118706
0.526315789473684 0.0821149318984471
1.05263157894737 0.0821149318984471
1.05263157894737 0.0813170961474685
1.57894736842105 0.0813170961474685
1.57894736842105 0.0811732322364271
2.10526315789474 0.0811732322364271
2.10526315789474 0.725485602761104
2.63157894736842 0.725485602761104
2.63157894736842 0.73816739723093
3.15789473684211 0.73816739723093
3.15789473684211 0.752847909773369
3.68421052631579 0.752847909773369
3.68421052631579 0.764164004417015
4.21052631578947 0.764164004417015
4.21052631578947 0.769265804438148
4.73684210526316 0.769265804438148
4.73684210526316 0.0974268657834384
5.26315789473684 0.0974268657834384
5.26315789473684 0.0970438729378922
5.78947368421053 0.0970438729378922
5.78947368421053 0.0965620532093126
6.31578947368421 0.0965620532093126
6.31578947368421 0.0960015095042879
6.84210526315789 0.0960015095042879
6.84210526315789 0.0953836231451416
7.36842105263158 0.0953836231451416
7.36842105263158 0.0947288812405935
7.89473684210526 0.0947288812405935
7.89473684210526 0.0940556565576741
8.42105263157895 0.0940556565576741
8.42105263157895 0.0933796580830246
8.94736842105263 0.0933796580830246
8.94736842105263 0.0927138092502769
9.47368421052632 0.0927138092502769
9.47368421052632 0.0920683793439983
10 0.0920683793439983
10 0.0914512403732916
11 0.0914512403732916
};
\addlegendentry{$\epsilon=1.7e-01\epsilon_c$, reuse}
\addplot [semithick, red, dashed]
table {%
0 0.0823042393748069
0.526315789473684 0.0823042393748069
0.526315789473684 0.0823042393674661
1.05263157894737 0.0823042393674661
1.05263157894737 0.0823042393608626
1.57894736842105 0.0823042393608626
1.57894736842105 0.0823042393586617
2.10526315789474 0.0823042393586617
2.10526315789474 0.746111735145988
2.63157894736842 0.746111735145988
2.63157894736842 0.746111735291274
3.15789473684211 0.746111735291274
3.15789473684211 0.746111735469007
3.68421052631579 0.746111735469007
3.68421052631579 0.746111735613752
4.21052631578947 0.746111735613752
4.21052631578947 0.746111735692488
4.73684210526316 0.746111735692488
4.73684210526316 0.0948252810887092
5.26315789473684 0.0948252810887092
5.26315789473684 0.094825281087524
5.78947368421053 0.094825281087524
5.78947368421053 0.0948252810852353
6.31578947368421 0.0948252810852353
6.31578947368421 0.0948252810819816
6.84210526315789 0.0948252810819816
6.84210526315789 0.0948252810779335
7.36842105263158 0.0948252810779335
7.36842105263158 0.0948252810732731
7.89473684210526 0.0948252810732731
7.89473684210526 0.0948252810682271
8.42105263157895 0.0948252810682271
8.42105263157895 0.0948252810630428
8.94736842105263 0.0948252810630428
8.94736842105263 0.0948252810580559
9.47368421052632 0.0948252810580559
9.47368421052632 0.0948252810537114
10 0.0948252810537114
10 0.0948252810505302
11 0.0948252810505302
};
\addlegendentry{$\epsilon=1.0e-05\ll\epsilon_c$, reuse}
\end{axis}

\end{tikzpicture}}
  \caption{The inversion result when the focusing parameter is slightly larger and equal to the critical value ($\epsilon_c$). For the latter case we consider both the start and reuse strategy. In the upper plot the Cauchy {measure} is used while the MGS {measure} is used in the bottom plot. Finally, we also plotted the conductivity profile when the focusing parameter is much smaller than the critical value.\label{fig:embed_critical} }
\end{figure}

Both strategies are illustrated in the flowchart of Figure~\ref{fig:diagram_start_reuse}. For every solution, we plot the data misfit in terms of the model misfit\footnote{We thus adopt a very similar strategy as in \citep{hansen1999curve} where a trade-off is made between data vs.~model misfit. This is different from the parameter selection in \citep{zhdanov2004minimum} where the model misfit is considered directly in terms of the focusing parameter itself, and the maximal curvature point on that curve is pinpointed to fix the parameter.}, see Figure~\ref{fig:embed_lcurve}.
With the Cauchy {measure}, we get a $J$-shaped curve which displays a jump in the data misfit on the horizontal line.
This discontinuity happens when decreasing $\epsilon$ from \num{1.56e-1} to \num{1.38e-1} and changes the model parameters from smooth to blocky (see Figure~\ref{fig:embed_critical}).
Using the Cauchy {measure}, one can find two interesting results: a blocky result just after the discontinuity and a smooth result for a focusing parameter larger than the critical value \num{1.38e-1} (see Figure~\ref{fig:embed_lcurve} (a)).
This critical value is selected as the optimal focusing parameter, producing a blocky result while still allowing for some smoothness. Even if for some specific surveys, a smoother result is expected, for one reason or another, this is also possible with the Cauchy {measure}.
In such case, the focusing parameter must then be larger than the critical value, but this still allows for a lot of freedom.
In our results we chose the value for which the data misfit is almost equal to the data misfit produced when the critical value is used.

In the case of the MGS {measure}, the curve is different, especially for smaller $\epsilon$.
For larger $\epsilon$, the model parameters are the same as with the Cauchy {measure} and we obtain a smooth shape.
The reuse method has two discontinuities, the first discontinuity, indicated with a star, changes the smooth profile to a semi-smooth profile with one jump.
The second discontinuity, indicated by a cross, transforms the profile to a blocky one.
With decreasing focusing parameters, the model parameters change only slightly.
If we use the start strategy, three discontinuities appear.
The first and last have the same effect as with the reuse strategy, while the second discontinuity, indicated by a square, changes the position of the jump in conductivity.
Note that the start strategy produces blocky model parameters with a much larger data misfit.

\begin{figure}
  \centering
  \subcaptionbox{\label{fig:embed_lcurve_cauchy}}{
\begin{tikzpicture}

\definecolor{darkgray176}{RGB}{176,176,176}
\definecolor{darkviolet1910191}{RGB}{191,0,191}
\definecolor{lightgray204}{RGB}{204,204,204}

\begin{axis}[
height=.5\textwidth,
legend cell align={left},
legend style={
  fill opacity=0.8,
  draw opacity=1,
  text opacity=1,
  at={(0.03,0.97)},
  anchor=north west,
  draw=lightgray204
},
log basis x={10},
log basis y={10},
minor xtick={1.6},
minor ytick={0.02,0.03,0.04,0.05,0.06,0.07,0.08,0.09,0.2,0.3,0.4,0.5,0.6,0.7,0.8,0.9,2,3,4,6,7,8,9,20,30,40,60,70,80,90,200,300,400,500,600,700,800,900,2000,3000,4000,5000,6000,7000,8000,9000},
tick align=outside,
tick pos=left,
title={Cauchy},
width=.5\textwidth,
x grid style={darkgray176},
xlabel={data misfit},
xmin=1, xmax=1.55520434543659,
xmode=log,
xtick style={color=black},
xtick={1,1.1,1.2,1.3,1.4,1.5},
y grid style={darkgray176},
ylabel={model misfit},
ymin=0.400827295826091, ymax=57.3994160571744,
ymode=log,
ytick style={color=black},
ytick={1,5,10,50}
]
\addplot [semithick, red, mark=triangle*, mark size=1, mark options={solid,rotate=180}, only marks]
table {%
1.52785483992261 45.1620024081989
1.52782765115606 44.6971050454869
1.52782757504387 44.2319370903373
1.52782752462705 43.7667688755151
1.52782751613206 43.3016002380474
1.52782751027013 42.8364315698658
1.52782750166937 42.3712629235337
1.52782749049367 41.9060942959627
1.52782748613369 41.4409255914289
1.52782747280902 40.9757569654096
1.52782745695794 40.5105883506174
1.5278274420982 40.0454197082034
1.52782741732366 39.5802511425795
1.52782738918404 39.1150825824044
1.52782734857173 38.6499141113596
1.52782729977628 38.1847456772376
1.52783304856402 37.719519240211
1.52783281647842 37.2543525094128
1.52783256406505 36.7891858916867
1.52785216649899 36.3238210459149
1.5278536373259 35.8586369985501
1.52785517948322 35.3934520645325
1.52783200863247 34.9285134075026
1.52783149726361 34.4633484225991
1.52785274733026 33.9979659622236
1.5278309423291 33.5330125929074
1.52783080186632 33.0678425012269
1.52784440288059 32.6025345375432
1.52782891297731 32.1375159997807
1.52782680645362 31.6723627356104
1.52782496073088 31.2072053981941
1.52782304497297 30.7420469081231
1.52782056555852 30.2768917127144
1.52781732675773 29.811741155571
1.52781335345635 29.346594215701
1.52780835411983 28.8814528315693
1.52780272926405 28.4163117718524
1.52779470980385 27.9511871608314
1.52778397594016 27.4860802360813
1.52777273833602 27.020966440139
1.5277549488508 26.5559030842971
1.5277351008884 26.0908413410098
1.5277094492506 25.6258136682393
1.52767720963894 25.1608216407439
1.52763856208665 24.6958555619514
1.52758731190577 24.2309673563106
1.52752251104625 23.7661539238436
1.52744051047706 23.3014358617006
1.52733747042589 22.8368315270849
1.527207457648 22.3723749637769
1.52704404509011 21.9080985595673
1.52683718594198 21.4440625578705
1.52657620619222 20.980323005088
1.52624700141829 20.5169570393552
1.52583172421742 20.0540626033981
1.52530812750001 19.5917607270179
1.52464837050511 19.130202095502
1.52381508412857 18.6695996519839
1.52276751199726 18.2101589056495
1.5214385127565 17.752297279472
1.51977038677484 17.2962733723661
1.51767519544734 16.8425675323359
1.51503606280186 16.3918497107801
1.51172403433415 15.9447876694019
1.50756076140257 15.5023924596401
1.50234470223159 15.0657246306835
1.49582842493423 14.6360856441343
1.48769526833683 14.2152143327818
1.47759811177752 13.804865447787
1.46512857881873 13.4070960613714
1.44984493698428 13.023993562474
1.43129702033867 12.6574891448699
1.40916514146192 12.3081284515526
1.38331670006944 11.9748138852217
1.35405865628065 11.6527763026269
1.32229339631493 11.3326654210125
1.28953932989114 11.000947931341
1.10258873810305 10.1710165100175
1.10236816422259 9.69581519458777
1.10219961623281 9.21673029738607
1.10213126116791 8.73221495829742
1.12530987003079 8.15813354507345
1.12258321410683 7.66248139697704
1.16145904207993 6.99250090195
1.14838314460926 6.25150830928385
1.13617949456479 5.57896074310129
1.12527273678691 4.96140829316187
1.11598650900501 4.38582570989836
1.10864723087946 3.83854494586582
1.1038312006809 3.30141154036515
1.10118298635614 2.76589669258232
1.09635044454256 2.30390594200499
1.09161581730387 1.91273228543642
1.08754489718499 1.58230976041112
1.08409693480312 1.30526261462102
1.08114045882353 1.07480916242098
1.07852815928281 0.884702551488682
1.07612160873326 0.729242488469854
1.07379953747719 0.603298292766985
1.07146259172223 0.502290569621078
};
\addlegendentry{Reuse}
\addplot [semithick, blue, mark=*, mark size=0.5, mark options={solid}, only marks]
table {%
1.15474834276938 45.8046678788939
1.15474834274936 45.3394991711189
1.15474834273713 44.8743304636547
1.15474834268077 44.4091617569264
1.15474834264423 43.9439930502039
1.15474834140853 43.4788243609732
1.15474834137754 43.0136556507317
1.1547483412347 42.5484869423559
1.15474834114198 42.0833182339951
1.15474833982359 41.618149538887
1.15474833967599 41.1529808289237
1.15474833757541 40.6878121400756
1.15474833734037 40.2226434275608
1.15474833569678 39.7574747289025
1.15474833362279 39.2923060320762
1.15474833100637 38.8271373375557
1.15474832772415 38.3619686457577
1.15474832001664 37.8968000014943
1.15474831825771 37.4316312749825
1.15474831162179 36.966462597581
1.1547483032486 36.5012939275741
1.15474829268329 36.0361252668978
1.15474827935179 35.5709566179951
1.15474826252984 35.1057879839489
1.15474824130339 34.6406193686493
1.15474821451915 34.1754507770052
1.15474818072176 33.710282215211
1.15474813807473 33.2451136910834
1.15474808426068 32.7799452144861
1.15474801635549 32.3147767978658
1.15474793066909 31.8496084569288
1.15474782254534 31.3844402114948
1.15474768610871 30.9192720865731
1.15474751394525 30.4541041137223
1.15474729669964 29.9889363327641
1.15474702256694 29.5237687939474
1.15474667665142 29.0586015606763
1.15474624015725 28.5934347129544
1.15474568936779 28.1282683517275
1.15474499435768 27.6631026043636
1.15474411736787 27.1979376315637
1.1547430107562 26.7327736360783
1.15474161441167 26.2676108737001
1.15473985249371 25.8024496671227
1.15473762931965 25.3372904234108
1.15473482417956 24.8721336560129
1.15473128479872 24.406980012503
1.15472681909855 23.9418303095019
1.15472118481722 23.4766855766177
1.15471407643101 23.0115471117635
1.15470510870059 22.5464165505663
1.15469379596815 22.0812959535197
1.15467952611065 21.616187915434
1.15466152788134 21.151095701864
1.15463882986748 20.6860234202237
1.15461020926478 20.2209762310856
1.15457412773166 19.755960611496
1.15452865109688 19.2909846832953
1.15447135094402 18.8260586024672
1.15439918112488 18.3611950498723
1.15430832735378 17.8964098064949
1.15419402318707 17.4317224376551
1.15405032789886 16.9671570807983
1.153869861755 16.5027433235035
1.15364349337875 16.0385171590725
1.15335991374464 15.5745226031181
1.15300403178528 15.1108245386051
1.15256193600505 14.6474553422848
1.15201249267268 14.1844944037173
1.15133231755345 13.7220133333692
1.15049445251956 13.2600834894409
1.14946870753375 12.798766415056
1.14822264245771 12.3380976973768
1.14672347675119 11.8780617224762
1.14493810921249 11.4185858338742
1.14284820853394 10.9593843514692
1.14044296651826 10.5000255569849
1.13773240482392 10.0397898130993
1.13476159556313 9.5775080866008
1.13161183331853 9.1115058865565
1.1284008896988 8.63950277614737
1.12530929035511 8.15813934080788
1.12258313515484 7.66248218635185
1.16146008416995 6.99249048486857
1.14838363741661 6.25150338610009
1.13617897468674 5.57896594215733
1.12527289351838 4.96140672462358
1.11598642348774 4.38582656439192
1.10864719591241 3.83854529543961
1.10383082692128 3.30141527636605
1.10118312946507 2.76589526268489
1.0963500716889 2.30390966627236
1.09161625087539 1.91272795385644
1.0875453933152 1.58230479729836
1.08409780981429 1.30525386016945
1.08114023669978 1.074811383675
1.07852754606686 0.88470868303049
1.07612160768976 0.729242498402535
1.07379953783645 0.603298289175159
1.07146203096808 0.502296177434523
};
\addlegendentry{Start}
\addplot [semithick, red, mark=asterisk, mark size=5, mark options={solid}, only marks, forget plot]
table {%
1.10213126116791 8.73221495829742
};
\addplot [semithick, blue, mark=asterisk, mark size=5, mark options={solid}, only marks, forget plot]
table {%
1.12258313515484 7.66248218635185
};
\addplot [semithick, darkviolet1910191, mark=asterisk, mark size=5, mark options={solid}, only marks, forget plot]
table {%
1.10118312946507 2.76589526268489
};
\end{axis}

\end{tikzpicture}}\subcaptionbox{}{\input{plots/embed_mgs}}
  \caption{(a) and (b) The data and model misfit of the different solvers. The data points with the label `start' used the first solution as a starting point, while the `reuse' approach uses the previous solution. The lower left corner corresponds with a larger focusing parameter. In Figure (b) the circle indicates the position of the last smooth model parameters. The star and the square show the critical value resulting in model parameters with one discontinuity. Finally, the cross is positioned at the critical value where the model parameters have two discontinuities.}\label{fig:embed_lcurve}
\end{figure}

\begin{figure}
  \centering
  \subcaptionbox{}{
\begin{tikzpicture}

\definecolor{darkgray176}{RGB}{176,176,176}
\definecolor{darkviolet1910191}{RGB}{191,0,191}
\definecolor{lightgray204}{RGB}{204,204,204}

\begin{axis}[
height=.3\textwidth,
legend cell align={left},
legend style={fill opacity=0.8, draw opacity=1, text opacity=1, draw=lightgray204},
log basis y={10},
tick align=outside,
tick pos=left,
width=.99\textwidth,
x grid style={darkgray176},
xlabel={d (m)},
xmin=-0.55, xmax=11.55,
xtick style={color=black},
y grid style={darkgray176},
ylabel={\(\displaystyle \sigma (\si{\milli\siemens\per\meter})\)},
ymin=0.03, ymax=1,
ymode=log,
ytick style={color=black},
ytick={0.05,0.1,0.5,1},
yticklabels={50,100,500,1000}
]
\addplot [very thick, black]
table {%
0 0.1
2.5 0.1
2.5 0.8
5 0.8
5 0.1
10 0.1
10 0.1
11 0.1
};
\addlegendentry{True}
\addplot [semithick, darkviolet1910191, dotted]
table {%
0 0.0878252899439021
0.526315789473684 0.0878252899439021
0.526315789473684 0.0824457794916471
1.05263157894737 0.0824457794916471
1.05263157894737 0.102884689001886
1.57894736842105 0.102884689001886
1.57894736842105 0.166276704110957
2.10526315789474 0.166276704110957
2.10526315789474 0.426883463830578
2.63157894736842 0.426883463830578
2.63157894736842 0.657639429090209
3.15789473684211 0.657639429090209
3.15789473684211 0.785776397524691
3.68421052631579 0.785776397524691
3.68421052631579 0.769024364745037
4.21052631578947 0.769024364745037
4.21052631578947 0.642722077349211
4.73684210526316 0.642722077349211
4.73684210526316 0.481412360066124
5.26315789473684 0.481412360066124
5.26315789473684 0.342375945566847
5.78947368421053 0.342375945566847
5.78947368421053 0.242423961809802
6.31578947368421 0.242423961809802
6.31578947368421 0.175040208425406
6.84210526315789 0.175040208425406
6.84210526315789 0.129805133298928
7.36842105263158 0.129805133298928
7.36842105263158 0.0988360600607397
7.89473684210526 0.0988360600607397
7.89473684210526 0.0770566043739039
8.42105263157895 0.0770566043739039
8.42105263157895 0.0613129891657968
8.94736842105263 0.0613129891657968
8.94736842105263 0.0496355855651163
9.47368421052632 0.0496355855651163
9.47368421052632 0.0407705391678196
10 0.0407705391678196
10 0.0338994274342457
11 0.0338994274342457
};
\addlegendentry{Smooth}
\addplot [semithick, blue, dashed]
table {%
0 0.087609117906306
0.526315789473684 0.087609117906306
0.526315789473684 0.0847620478421255
1.05263157894737 0.0847620478421255
1.05263157894737 0.0859320502344069
1.57894736842105 0.0859320502344069
1.57894736842105 0.0895671508450695
2.10526315789474 0.0895671508450695
2.10526315789474 0.629412077567342
2.63157894736842 0.629412077567342
2.63157894736842 0.672045061646295
3.15789473684211 0.672045061646295
3.15789473684211 0.706254779950321
3.68421052631579 0.706254779950321
3.68421052631579 0.722402738529109
4.21052631578947 0.722402738529109
4.21052631578947 0.718152266567957
4.73684210526316 0.718152266567957
4.73684210526316 0.696002237208224
5.26315789473684 0.696002237208224
5.26315789473684 0.0955898643858094
5.78947368421053 0.0955898643858094
5.78947368421053 0.0906397397185429
6.31578947368421 0.0906397397185429
6.31578947368421 0.0858696662034068
6.84210526315789 0.0858696662034068
6.84210526315789 0.081354960012311
7.36842105263158 0.081354960012311
7.36842105263158 0.0771407680137099
7.89473684210526 0.0771407680137099
7.89473684210526 0.0732478263505798
8.42105263157895 0.0732478263505798
8.42105263157895 0.0696788656514016
8.94736842105263 0.0696788656514016
8.94736842105263 0.0664244101427341
9.47368421052632 0.0664244101427341
9.47368421052632 0.0634674458937306
10 0.0634674458937306
10 0.0607868746106451
11 0.0607868746106451
};
\addlegendentry{Blocky Start}
\addplot [semithick, red, dash pattern=on 1pt off 3pt on 3pt off 3pt]
table {%
0 0.0878741527397826
0.526315789473684 0.0878741527397826
0.526315789473684 0.0869764979990246
1.05263157894737 0.0869764979990246
1.05263157894737 0.0884027874781238
1.57894736842105 0.0884027874781238
1.57894736842105 0.0910581726012711
2.10526315789474 0.0910581726012711
2.10526315789474 0.628336811345192
2.63157894736842 0.628336811345192
2.63157894736842 0.647128639694423
3.15789473684211 0.647128639694423
3.15789473684211 0.65812430714756
3.68421052631579 0.65812430714756
3.68421052631579 0.660001035825451
4.21052631578947 0.660001035825451
4.21052631578947 0.653794185982952
4.73684210526316 0.653794185982952
4.73684210526316 0.641671856423061
5.26315789473684 0.641671856423061
5.26315789473684 0.626195499897184
5.78947368421053 0.626195499897184
5.78947368421053 0.0562160713868449
6.31578947368421 0.0562160713868449
6.31578947368421 0.0547507650704529
6.84210526315789 0.0547507650704529
6.84210526315789 0.0533422008902618
7.36842105263158 0.0533422008902618
7.36842105263158 0.0519996793713607
7.89473684210526 0.0519996793713607
7.89473684210526 0.0507289885372542
8.42105263157895 0.0507289885372542
8.42105263157895 0.0495331183588796
8.94736842105263 0.0495331183588796
8.94736842105263 0.048412863653932
9.47368421052632 0.048412863653932
9.47368421052632 0.0473673685636454
10 0.0473673685636454
10 0.046394619171991
11 0.046394619171991
};
\addlegendentry{Blocky Reuse}
\end{axis}

\end{tikzpicture}}
  \subcaptionbox{}{
\begin{tikzpicture}

\definecolor{darkgray176}{RGB}{176,176,176}
\definecolor{darkviolet1910191}{RGB}{191,0,191}
\definecolor{green01270}{RGB}{0,127,0}
\definecolor{lightgray204}{RGB}{204,204,204}
\begin{axis}[
height=.3\textwidth,
legend cell align={left},
legend style={fill opacity=0.8, draw opacity=1, text opacity=1, draw=lightgray204},
log basis y={10},
tick align=outside,
tick pos=left,
width=.99\textwidth,
x grid style={darkgray176},
xlabel={d (m)},
xmin=-0.55, xmax=11.55,
xtick style={color=black},
y grid style={darkgray176},
ylabel={\(\displaystyle \sigma (\si{\milli\siemens\per\meter})\)},
ymin=0.05, ymax=1,
ymode=log,
ytick style={color=black},
ytick={0.05,0.1,0.5,1},
yticklabels={50,100,500,1000}
  ]
\addplot [very thick, black]
table {%
0 0.1
2.5 0.1
2.5 0.8
5 0.8
5 0.1
10 0.1
10 0.1
11 0.1
};
\addlegendentry{True}
\addplot [semithick, darkviolet1910191, dotted]
table {%
0 0.0817929032649012
0.526315789473684 0.0817929032649012
0.526315789473684 0.0927646222130712
1.05263157894737 0.0927646222130712
1.05263157894737 0.129527238788964
1.57894736842105 0.129527238788964
1.57894736842105 0.227693824611986
2.10526315789474 0.227693824611986
2.10526315789474 0.430985290059666
2.63157894736842 0.430985290059666
2.63157894736842 0.604245057817819
3.15789473684211 0.604245057817819
3.15789473684211 0.693577892143219
3.68421052631579 0.693577892143219
3.68421052631579 0.676158414374681
4.21052631578947 0.676158414374681
4.21052631578947 0.580982059449739
4.73684210526316 0.580982059449739
4.73684210526316 0.459312360485411
5.26315789473684 0.459312360485411
5.26315789473684 0.349204043845573
5.78947368421053 0.349204043845573
5.78947368421053 0.264071203362216
6.31578947368421 0.264071203362216
6.31578947368421 0.202268805080891
6.84210526315789 0.202268805080891
6.84210526315789 0.15800783518227
7.36842105263158 0.15800783518227
7.36842105263158 0.126022377403446
7.89473684210526 0.126022377403446
7.89473684210526 0.102489516128801
8.42105263157895 0.102489516128801
8.42105263157895 0.0848147789180583
8.94736842105263 0.0848147789180583
8.94736842105263 0.0712631970545012
9.47368421052632 0.0712631970545012
9.47368421052632 0.0606686546075611
10 0.0606686546075611
10 0.05223521494328
11 0.05223521494328
};
\addlegendentry{Smooth}
\addplot [semithick, red]
table {%
0 0.0859052518081656
0.526315789473684 0.0859052518081656
0.526315789473684 0.0851805117353463
1.05263157894737 0.0851805117353463
1.05263157894737 0.0900902191066861
1.57894736842105 0.0900902191066861
1.57894736842105 0.0982095462745568
2.10526315789474 0.0982095462745568
2.10526315789474 0.706461724453108
2.63157894736842 0.706461724453108
2.63157894736842 0.724557747226347
3.15789473684211 0.724557747226347
3.15789473684211 0.681334446651552
3.68421052631579 0.681334446651552
3.68421052631579 0.591771384040648
4.21052631578947 0.591771384040648
4.21052631578947 0.48501878337905
4.73684210526316 0.48501878337905
4.73684210526316 0.385145701797884
5.26315789473684 0.385145701797884
5.26315789473684 0.303213677228051
5.78947368421053 0.303213677228051
5.78947368421053 0.240226670300209
6.31578947368421 0.240226670300209
6.31578947368421 0.192957567759496
6.84210526315789 0.192957567759496
6.84210526315789 0.157544645719435
7.36842105263158 0.157544645719435
7.36842105263158 0.130766580439604
7.89473684210526 0.130766580439604
7.89473684210526 0.110230564004458
8.42105263157895 0.110230564004458
8.42105263157895 0.0942314150641939
8.94736842105263 0.0942314150641939
8.94736842105263 0.0815671277968088
9.47368421052632 0.0815671277968088
9.47368421052632 0.0713878685273951
10 0.0713878685273951
10 0.0630871182767593
11 0.0630871182767593
};
\addlegendentry{Semi-smooth}
\addplot [semithick, blue, dash pattern=on 1pt off 3pt on 3pt off 3pt]
table {%
0 0.0772458582366291
0.526315789473684 0.0772458582366291
0.526315789473684 0.0736680419712923
1.05263157894737 0.0736680419712923
1.05263157894737 0.0738845661736059
1.57894736842105 0.0738845661736059
1.57894736842105 0.484002620193582
2.10526315789474 0.484002620193582
2.10526315789474 0.5228335882311
2.63157894736842 0.5228335882311
2.63157894736842 0.549919709327459
3.15789473684211 0.549919709327459
3.15789473684211 0.546672849090581
3.68421052631579 0.546672849090581
3.68421052631579 0.51147036979972
4.21052631578947 0.51147036979972
4.21052631578947 0.453803315842151
4.73684210526316 0.453803315842151
4.73684210526316 0.387177458218394
5.26315789473684 0.387177458218394
5.26315789473684 0.323037251732601
5.78947368421053 0.323037251732601
5.78947368421053 0.267687073528505
6.31578947368421 0.267687073528505
6.31578947368421 0.222720482729021
6.84210526315789 0.222720482729021
6.84210526315789 0.187160185124095
7.36842105263158 0.187160185124095
7.36842105263158 0.159231329965352
7.89473684210526 0.159231329965352
7.89473684210526 0.137212789045507
8.42105263157895 0.137212789045507
8.42105263157895 0.119697619814678
8.94736842105263 0.119697619814678
8.94736842105263 0.105608306831528
9.47368421052632 0.105608306831528
9.47368421052632 0.0941395844264112
10 0.0941395844264112
10 0.0846932856698104
11 0.0846932856698104
};
\addlegendentry{Alt. semi-smooth}
\addplot [semithick, green01270, dashed]
table {%
0 0.072671149716147
0.526315789473684 0.072671149716147
0.526315789473684 0.0715742701826133
1.05263157894737 0.0715742701826133
1.05263157894737 0.0711388847631433
1.57894736842105 0.0711388847631433
1.57894736842105 0.510554166609597
2.10526315789474 0.510554166609597
2.10526315789474 0.523421928265589
2.63157894736842 0.523421928265589
2.63157894736842 0.540961204653078
3.15789473684211 0.540961204653078
3.15789473684211 0.558126724767106
3.68421052631579 0.558126724767106
3.68421052631579 0.571941736037987
4.21052631578947 0.571941736037987
4.21052631578947 0.581010569878072
4.73684210526316 0.581010569878072
4.73684210526316 0.585004697126048
5.26315789473684 0.585004697126048
5.26315789473684 0.10824063783668
5.78947368421053 0.10824063783668
5.78947368421053 0.107962007642161
6.31578947368421 0.107962007642161
6.31578947368421 0.10757159288735
6.84210526315789 0.10757159288735
6.84210526315789 0.107097849922875
7.36842105263158 0.107097849922875
7.36842105263158 0.106567031286516
7.89473684210526 0.106567031286516
7.89473684210526 0.106002113617231
8.42105263157895 0.106002113617231
8.42105263157895 0.10542242312848
8.94736842105263 0.10542242312848
8.94736842105263 0.10484369256431
9.47368421052632 0.10484369256431
9.47368421052632 0.104278354674583
10 0.104278354674583
10 0.103735933932057
11 0.103735933932057
};
\addlegendentry{Blocky}
\end{axis}

\end{tikzpicture}}
  \caption{(a) The model parameters when using the Cauchy measure. The focusing parameters that are used in the inverse problem are indicated with a star in Figure~\ref{fig:embed_lcurve_cauchy}. (b) The model parameters of the different regions as shown in the upper right plot of Figure \ref{fig:embed_lcurve} (b) when using the MGS measure. The alternative semi-smooth region is only present with the start strategy. To prevent cluttering the figure, the optimal blocky result with the reuse strategy is not plotted. }\label{fig:embed_lcurve_layers}
\end{figure}

In Figure~\ref{fig:embed_lcurve} (b) the regions of the different model parameters are plotted if the MGS {measure} is used.
The smooth and semi-smooth regions, which both strategies have in common, fail to find the conductivity of the third layer and the second interface depth.
The semi-smooth result, however, finds the first interface depth.
With the start strategy, we get another alternative semi-smooth region after the second discontinuity.
The data misfit has increased significantly (from 1.2 to 1.6), and the model parameters have a smaller peak value than the results from the smooth and semi-smooth regions.
After the third discontinuity, the second one for the reuse strategy, we find the blocky region.
This result shows the two interface depth clearly, with the reuse strategy finding almost the correct width (\SI{2.5}{\meter}) and height (\SI{700}{\milli\siemens\per\meter}).
With the start strategy, the width (\SI{4}{\meter}) and height (\SI{500}{\milli\siemens\per\meter}) of the peak deviates more from the true values, again resulting in a much larger data misfit.  We have shown the corresponding model parameters per measure in Figure 5 (a) and (b).

\section{Results}
\label{sec:results}

To illustrate the performance of our algorithm we test our method on a synthetic conductivity profile, a profile based on borehole logging and data collected at the Belgian coast.
In all subsequent figures, we show three inversion results for both the Cauchy and MGS {measure}.
 {We did not use a separate $\ell_2$-inversion procedure to get to the shown smooth conductivity profile. Rather it is obtained from an inversion with a focusing parameter which is sufficiently larger than the critical value ($\epsilon=0.50$ and $\epsilon=0.63$ for the Cauchy and MGS measure respectively), in which case we effectively fall back to an $\ell_2$-regularisation, cf.~the discussion below Eq.~\eqref{normCauchy}, while we can use the same numerical routine as for the other inversions. }

  The focusing parameter was chosen so that the data misfit was almost equal to the data misfit of the other results.
  For the smooth results we do not distinguish between the start and reuse strategy, as both generate similar profiles.

The profiles labeled as start and reuse were obtained with the corresponding strategy.
  The selected focusing parameter is equal to the critical value, that is the first focusing parameter for which a blocky profile is obtained.
  This can not only be seen from the profile itself, but also from a sudden jump in the data misfit.

\subsection{Inversion of synthetic data}
\label{sec:synth_data}
\begin{figure}[t]
  \centering
  \subcaptionbox{}{
\begin{tikzpicture}

\definecolor{color0}{rgb}{0.75,0,0.75}

\begin{axis}[
height=.3\textwidth,
legend cell align={left},
legend style={
  fill opacity=0.8,
  draw opacity=1,
  text opacity=1,
  at={(0.97,0.03)},
  anchor=south east,
  draw=white!80!black
},
log basis y={10},
tick align=outside,
tick pos=left,
title={Cauchy},
width=.5\textwidth,
x grid style={white!69.0196078431373!black},
xlabel={d (m)},
xmin=-0.55, xmax=11.55,
xtick style={color=black},
y grid style={white!69.0196078431373!black},
ylabel={\(\displaystyle \sigma\) (Sm\(\displaystyle ^-1\))},
ymin=0.007, ymax=0.13,
ymode=log,
ytick style={color=black}
]
\addplot [semithick, red]
table {%
0 0.00980142599774755
0.526315789473684 0.00980142599774755
0.526315789473684 0.00996342643760328
1.05263157894737 0.00996342643760328
1.05263157894737 0.0102731013339251
1.57894736842105 0.0102731013339251
1.57894736842105 0.0105405506465555
2.10526315789474 0.0105405506465555
2.10526315789474 0.122670106677196
2.63157894736842 0.122670106677196
2.63157894736842 0.116028872579617
3.15789473684211 0.116028872579617
3.15789473684211 0.0749710155720998
3.68421052631579 0.0749710155720998
3.68421052631579 0.0685391410342466
4.21052631578947 0.0685391410342466
4.21052631578947 0.0635702062845243
4.73684210526316 0.0635702062845243
4.73684210526316 0.0598443486204936
5.26315789473684 0.0598443486204936
5.26315789473684 0.0570303395028081
5.78947368421053 0.0570303395028081
5.78947368421053 0.0548605296651888
6.31578947368421 0.0548605296651888
6.31578947368421 0.0531387213099956
6.84210526315789 0.0531387213099956
6.84210526315789 0.0517238411929323
7.36842105263158 0.0517238411929323
7.36842105263158 0.0505150605941921
7.89473684210526 0.0505150605941921
7.89473684210526 0.0494409601570395
8.42105263157895 0.0494409601570395
8.42105263157895 0.0484518024730905
8.94736842105263 0.0484518024730905
8.94736842105263 0.0475138916280952
9.47368421052632 0.0475138916280952
9.47368421052632 0.0466053425815034
10 0.0466053425815034
10 0.045712850039654
11 0.045712850039654
};
\addlegendentry{Reuse}
\addplot [semithick, blue]
table {%
0 0.00986612738822728
0.526315789473684 0.00986612738822728
0.526315789473684 0.00991906469316734
1.05263157894737 0.00991906469316734
1.05263157894737 0.00997942670436207
1.57894736842105 0.00997942670436207
1.57894736842105 0.0100264586551598
2.10526315789474 0.0100264586551598
2.10526315789474 0.115939041276795
2.63157894736842 0.115939041276795
2.63157894736842 0.115200345647112
3.15789473684211 0.115200345647112
3.15789473684211 0.114052842950327
3.68421052631579 0.114052842950327
3.68421052631579 0.0538705679795384
4.21052631578947 0.0538705679795384
4.21052631578947 0.0533221753122781
4.73684210526316 0.0533221753122781
4.73684210526316 0.0527919308076837
5.26315789473684 0.0527919308076837
5.26315789473684 0.052267775541649
5.78947368421053 0.052267775541649
5.78947368421053 0.0517367514728307
6.31578947368421 0.0517367514728307
6.31578947368421 0.0511888211702565
6.84210526315789 0.0511888211702565
6.84210526315789 0.050617930611599
7.36842105263158 0.050617930611599
7.36842105263158 0.050021786269429
7.89473684210526 0.050021786269429
7.89473684210526 0.049401125188381
8.42105263157895 0.049401125188381
8.42105263157895 0.0487588641027876
8.94736842105263 0.0487588641027876
8.94736842105263 0.0480993068928516
9.47368421052632 0.0480993068928516
9.47368421052632 0.0474274833406979
10 0.0474274833406979
10 0.0467486389785805
11 0.0467486389785805
};
\addlegendentry{Start}
\addplot [semithick, color0]
table {%
0 0.00998675423598189
0.526315789473684 0.00998675423598189
0.526315789473684 0.00848647243317177
1.05263157894737 0.00848647243317177
1.05263157894737 0.0107816699113292
1.57894736842105 0.0107816699113292
1.57894736842105 0.0179603367167187
2.10526315789474 0.0179603367167187
2.10526315789474 0.117425396441199
2.63157894736842 0.117425396441199
2.63157894736842 0.11038511256927
3.15789473684211 0.11038511256927
3.15789473684211 0.0851536446847102
3.68421052631579 0.0851536446847102
3.68421052631579 0.0667455339028673
4.21052631578947 0.0667455339028673
4.21052631578947 0.0567950968521003
4.73684210526316 0.0567950968521003
4.73684210526316 0.0522331304114151
5.26315789473684 0.0522331304114151
5.26315789473684 0.0510126464053439
5.78947368421053 0.0510126464053439
5.78947368421053 0.0519175617760298
6.31578947368421 0.0519175617760298
6.31578947368421 0.0540572980538688
6.84210526315789 0.0540572980538688
6.84210526315789 0.056586244038746
7.36842105263158 0.056586244038746
7.36842105263158 0.0586277322803191
7.89473684210526 0.0586277322803191
7.89473684210526 0.0593686274266818
8.42105263157895 0.0593686274266818
8.42105263157895 0.0582487565279
8.94736842105263 0.0582487565279
8.94736842105263 0.0551201441232352
9.47368421052632 0.0551201441232352
9.47368421052632 0.0502691633622223
10 0.0502691633622223
10 0.0442848149237943
11 0.0442848149237943
};
\addlegendentry{Smooth}
\addplot [very thick, black]
table {%
0 0.01
2 0.01
2 0.1
4 0.1
4 0.05
5 0.05
};
\addlegendentry{True}
\end{axis}

\end{tikzpicture}}\subcaptionbox{}{
\begin{tikzpicture}

\definecolor{color0}{rgb}{0.75,0,0.75}

\begin{axis}[
height=.3\textwidth,
legend cell align={left},
legend style={
  fill opacity=0.8,
  draw opacity=1,
  text opacity=1,
  at={(0.97,0.03)},
  anchor=south east,
  draw=white!80!black
},
log basis y={10},
tick align=outside,
tick pos=left,
title={MGS},
width=.5\textwidth,
x grid style={white!69.0196078431373!black},
xlabel={d (m)},
xmin=-0.55, xmax=11.55,
xtick style={color=black},
y grid style={white!69.0196078431373!black},
ylabel={\(\displaystyle \sigma\) (Sm\(\displaystyle ^-1\))},
ymin=0.007, ymax=0.13,
ymode=log,
ytick style={color=black}
]
\addplot [semithick, red]
table {%
0 0.00906087465790217
0.526315789473684 0.00906087465790217
0.526315789473684 0.00900493612270109
1.05263157894737 0.00900493612270109
1.05263157894737 0.00898357597255983
1.57894736842105 0.00898357597255983
1.57894736842105 0.0693885117172298
2.10526315789474 0.0693885117172298
2.10526315789474 0.0699067715340558
2.63157894736842 0.0699067715340558
2.63157894736842 0.0705710725601603
3.15789473684211 0.0705710725601603
3.15789473684211 0.0712064741480888
3.68421052631579 0.0712064741480888
3.68421052631579 0.0717453650480753
4.21052631578947 0.0717453650480753
4.21052631578947 0.072171367483974
4.73684210526316 0.072171367483974
4.73684210526316 0.0724895378442505
5.26315789473684 0.0724895378442505
5.26315789473684 0.0727127056075339
5.78947368421053 0.0727127056075339
5.78947368421053 0.0728555563705046
6.31578947368421 0.0728555563705046
6.31578947368421 0.0729321704346667
6.84210526315789 0.0729321704346667
6.84210526315789 0.0729550960068017
7.36842105263158 0.0729550960068017
7.36842105263158 0.0436456304145398
7.89473684210526 0.0436456304145398
7.89473684210526 0.0436215408417451
8.42105263157895 0.0436215408417451
8.42105263157895 0.0435880092928402
8.94736842105263 0.0435880092928402
8.94736842105263 0.0435472413112172
9.47368421052632 0.0435472413112172
9.47368421052632 0.0435010388144977
10 0.0435010388144977
10 0.043450868765097
11 0.043450868765097
};
\addlegendentry{Reuse}
\addplot [semithick, blue]
table {%
0 0.00997143420603288
0.526315789473684 0.00997143420603288
0.526315789473684 0.0100307403168109
1.05263157894737 0.0100307403168109
1.05263157894737 0.0100669188407222
1.57894736842105 0.0100669188407222
1.57894736842105 0.0100802076545704
2.10526315789474 0.0100802076545704
2.10526315789474 0.108778603987195
2.63157894736842 0.108778603987195
2.63157894736842 0.108208893940828
3.15789473684211 0.108208893940828
3.15789473684211 0.107614247188513
3.68421052631579 0.107614247188513
3.68421052631579 0.107245138414909
4.21052631578947 0.107245138414909
4.21052631578947 0.0457925639873932
4.73684210526316 0.0457925639873932
4.73684210526316 0.0458190206105692
5.26315789473684 0.0458190206105692
5.26315789473684 0.0458923889455197
5.78947368421053 0.0458923889455197
5.78947368421053 0.0460029360943722
6.31578947368421 0.0460029360943722
6.31578947368421 0.046140976270168
6.84210526315789 0.046140976270168
6.84210526315789 0.0462978562494868
7.36842105263158 0.0462978562494868
7.36842105263158 0.0464662998047138
7.89473684210526 0.0464662998047138
7.89473684210526 0.0466404364700202
8.42105263157895 0.0466404364700202
8.42105263157895 0.0468156850887041
8.94736842105263 0.0468156850887041
8.94736842105263 0.046988579555097
9.47368421052632 0.046988579555097
9.47368421052632 0.047156582732568
10 0.047156582732568
10 0.0473179108534462
11 0.0473179108534462
};
\addlegendentry{Start}
\addplot [semithick, color0]
table {%
0 0.00980472808612288
0.526315789473684 0.00980472808612288
0.526315789473684 0.00812237206709955
1.05263157894737 0.00812237206709955
1.05263157894737 0.00846470410125792
1.57894736842105 0.00846470410125792
1.57894736842105 0.054940861049638
2.10526315789474 0.054940861049638
2.10526315789474 0.0716268166183707
2.63157894736842 0.0716268166183707
2.63157894736842 0.0846075549310645
3.15789473684211 0.0846075549310645
3.15789473684211 0.0881467296115804
3.68421052631579 0.0881467296115804
3.68421052631579 0.0841345064255506
4.21052631578947 0.0841345064255506
4.21052631578947 0.0768263516159971
4.73684210526316 0.0768263516159971
4.73684210526316 0.0692931470692849
5.26315789473684 0.0692931470692849
5.26315789473684 0.062842455032783
5.78947368421053 0.062842455032783
5.78947368421053 0.0577436045156029
6.31578947368421 0.0577436045156029
6.31578947368421 0.0538642309583388
6.84210526315789 0.0538642309583388
6.84210526315789 0.0509743541155565
7.36842105263158 0.0509743541155565
7.36842105263158 0.04885461590149
7.89473684210526 0.04885461590149
7.89473684210526 0.0473240123167288
8.42105263157895 0.0473240123167288
8.42105263157895 0.0462404305159083
8.94736842105263 0.0462404305159083
8.94736842105263 0.0454940428307165
9.47368421052632 0.0454940428307165
9.47368421052632 0.0450001545212182
10 0.0450001545212182
10 0.044693271454481
11 0.044693271454481
};
\addlegendentry{Smooth}
\addplot [very thick, black]
table {%
0 0.01
2 0.01
2 0.1
4 0.1
4 0.05
5 0.05
};
\addlegendentry{True}
\end{axis}

\end{tikzpicture}}
  \subcaptionbox{}{
\begin{tikzpicture}

\definecolor{color0}{rgb}{0.75,0,0.75}

\begin{axis}[
height=.3\textwidth,
legend cell align={left},
legend style={
  fill opacity=0.8,
  draw opacity=1,
  text opacity=1,
  at={(0.97,0.03)},
  anchor=south east,
  draw=white!80!black
},
log basis y={10},
tick align=outside,
tick pos=left,
width=.5\textwidth,
x grid style={white!69.0196078431373!black},
xlabel={d (m)},
xmin=-0.55, xmax=11.55,
xtick style={color=black},
y grid style={white!69.0196078431373!black},
ylabel={\(\displaystyle \sigma\) (Sm\(\displaystyle ^-1\))},
ymin=0.007, ymax=0.12,
ymode=log,
ytick style={color=black}
]
\addplot [semithick, red]
table {%
0 0.0102366413328393
0.526315789473684 0.0102366413328393
0.526315789473684 0.010253755790571
1.05263157894737 0.010253755790571
1.05263157894737 0.01026949928511
1.57894736842105 0.01026949928511
1.57894736842105 0.0794378621935682
2.10526315789474 0.0794378621935682
2.10526315789474 0.0793563427790572
2.63157894736842 0.0793563427790572
2.63157894736842 0.0791429922099688
3.15789473684211 0.0791429922099688
3.15789473684211 0.0788366108889886
3.68421052631579 0.0788366108889886
3.68421052631579 0.0784669471705539
4.21052631578947 0.0784669471705539
4.21052631578947 0.0780563140841545
4.73684210526316 0.0780563140841545
4.73684210526316 0.0502458994444384
5.26315789473684 0.0502458994444384
5.26315789473684 0.0499600987766618
5.78947368421053 0.0499600987766618
5.78947368421053 0.0496739782584713
6.31578947368421 0.0496739782584713
6.31578947368421 0.0493905698023416
6.84210526315789 0.0493905698023416
6.84210526315789 0.0491121344834836
7.36842105263158 0.0491121344834836
7.36842105263158 0.0488403391704787
7.89473684210526 0.0488403391704787
7.89473684210526 0.0485763849377443
8.42105263157895 0.0485763849377443
8.42105263157895 0.0483211063475731
8.94736842105263 0.0483211063475731
8.94736842105263 0.0480750498438262
9.47368421052632 0.0480750498438262
9.47368421052632 0.0478385362851767
10 0.0478385362851767
10 0.0476117113404742
11 0.0476117113404742
};
\addlegendentry{Reuse}
\addplot [semithick, blue]
table {%
0 0.00996980219224632
0.526315789473684 0.00996980219224632
0.526315789473684 0.00997371141129148
1.05263157894737 0.00997371141129148
1.05263157894737 0.00998168432222106
1.57894736842105 0.00998168432222106
1.57894736842105 0.0845958343010148
2.10526315789474 0.0845958343010148
2.10526315789474 0.0846154490460769
2.63157894736842 0.0846154490460769
2.63157894736842 0.084517567657952
3.15789473684211 0.084517567657952
3.15789473684211 0.0842855339378233
3.68421052631579 0.0842855339378233
3.68421052631579 0.05352426276074
4.21052631578947 0.05352426276074
4.21052631578947 0.0532481021608462
4.73684210526316 0.0532481021608462
4.73684210526316 0.0529345141566993
5.26315789473684 0.0529345141566993
5.26315789473684 0.0525922077159632
5.78947368421053 0.0525922077159632
5.78947368421053 0.0522294400150782
6.31578947368421 0.0522294400150782
6.31578947368421 0.0518536013140156
6.84210526315789 0.0518536013140156
6.84210526315789 0.0514710324913725
7.36842105263158 0.0514710324913725
7.36842105263158 0.0510869909280399
7.89473684210526 0.0510869909280399
7.89473684210526 0.050705706259555
8.42105263157895 0.050705706259555
8.42105263157895 0.0503304849646342
8.94736842105263 0.0503304849646342
8.94736842105263 0.0499638357947164
9.47368421052632 0.0499638357947164
9.47368421052632 0.0496075981103417
10 0.0496075981103417
10 0.0492630625023522
11 0.0492630625023522
};
\addlegendentry{Start}
\addplot [semithick, color0]
table {%
0 0.00940228042821305
0.526315789473684 0.00940228042821305
0.526315789473684 0.012345754512616
1.05263157894737 0.012345754512616
1.05263157894737 0.0224277803100709
1.57894736842105 0.0224277803100709
1.57894736842105 0.0548698786540621
2.10526315789474 0.0548698786540621
2.10526315789474 0.0755270216532872
2.63157894736842 0.0755270216532872
2.63157894736842 0.0846412937656856
3.15789473684211 0.0846412937656856
3.15789473684211 0.083949170474504
3.68421052631579 0.083949170474504
3.68421052631579 0.0782350646414422
4.21052631578947 0.0782350646414422
4.21052631578947 0.0713352538398666
4.73684210526316 0.0713352538398666
4.73684210526316 0.0650751132478607
5.26315789473684 0.0650751132478607
5.26315789473684 0.0599762120107138
5.78947368421053 0.0599762120107138
5.78947368421053 0.0560127581431948
6.31578947368421 0.0560127581431948
6.31578947368421 0.0529976934122291
6.84210526315789 0.0529976934122291
6.84210526315789 0.0507289810540223
7.36842105263158 0.0507289810540223
7.36842105263158 0.0490330076922279
7.89473684210526 0.0490330076922279
7.89473684210526 0.0477713851610249
8.42105263157895 0.0477713851610249
8.42105263157895 0.0468368284867785
8.94736842105263 0.0468368284867785
8.94736842105263 0.046146951770264
9.47368421052632 0.046146951770264
9.47368421052632 0.0456386430448141
10 0.0456386430448141
10 0.045263578820726
11 0.045263578820726
};
\addlegendentry{Smooth}
\addplot [very thick, black]
table {%
0 0.01
1 0.01
1 0.0183890990440511
1.2 0.0183890990440511
1.2 0.0268179604518561
1.4 0.0268179604518561
1.4 0.0355174882966508
1.6 0.0355174882966508
1.6 0.0443816010632889
1.8 0.0443816010632889
1.8 0.0536834099310765
2 0.0536834099310765
2 0.0626555108972291
2.2 0.0626555108972291
2.2 0.0713084110276481
2.4 0.0713084110276481
2.4 0.0795286510250701
2.6 0.0795286510250701
2.6 0.0888777554465577
2.8 0.0888777554465577
2.8 0.0967470479737341
3 0.0967470479737341
3 0.1
3.2 0.1
3.2 0.0908223478234653
3.4 0.0908223478234653
3.4 0.0854014900439411
3.6 0.0854014900439411
3.6 0.0813074399360079
3.8 0.0813074399360079
3.8 0.0759351570996025
4 0.0759351570996025
4 0.0703729891292078
4.2 0.0703729891292078
4.2 0.065787352108943
4.4 0.065787352108943
4.4 0.0606915714904508
4.6 0.0606915714904508
4.6 0.0555576667684278
4.8 0.0555576667684278
4.8 0.05010310142021
5 0.05010310142021
5 0.05
6 0.05
};
\addlegendentry{True}
\end{axis}

\end{tikzpicture}}\subcaptionbox{}{
\begin{tikzpicture}

\definecolor{color0}{rgb}{0.75,0,0.75}

\begin{axis}[
height=.3\textwidth,
legend cell align={left},
legend style={
  fill opacity=0.8,
  draw opacity=1,
  text opacity=1,
  at={(0.97,0.03)},
  anchor=south east,
  draw=white!80!black
},
log basis y={10},
tick align=outside,
tick pos=left,
width=.5\textwidth,
x grid style={white!69.0196078431373!black},
xlabel={d (m)},
xmin=-0.55, xmax=11.55,
xtick style={color=black},
y grid style={white!69.0196078431373!black},
ylabel={\(\displaystyle \sigma\) (Sm\(\displaystyle ^-1\))},
ymin=0.007, ymax=0.12,
ymode=log,
ytick style={color=black}
]
\addplot [semithick, red]
table {%
0 0.0100232363154783
0.526315789473684 0.0100232363154783
0.526315789473684 0.0100104730106901
1.05263157894737 0.0100104730106901
1.05263157894737 0.0100043061473398
1.57894736842105 0.0100043061473398
1.57894736842105 0.0817875704427288
2.10526315789474 0.0817875704427288
2.10526315789474 0.0819852065024768
2.63157894736842 0.0819852065024768
2.63157894736842 0.0822266177070595
3.15789473684211 0.0822266177070595
3.15789473684211 0.0824173279224714
3.68421052631579 0.0824173279224714
3.68421052631579 0.0825041880616578
4.21052631578947 0.0825041880616578
4.21052631578947 0.0508117354579321
4.73684210526316 0.0508117354579321
4.73684210526316 0.0507344055207812
5.26315789473684 0.0507344055207812
5.26315789473684 0.0506075922390451
5.78947368421053 0.0506075922390451
5.78947368421053 0.0504358757208112
6.31578947368421 0.0504358757208112
6.31578947368421 0.0502251956999383
6.84210526315789 0.0502251956999383
6.84210526315789 0.0499820163869832
7.36842105263158 0.0499820163869832
7.36842105263158 0.0497127760571253
7.89473684210526 0.0497127760571253
7.89473684210526 0.0494235449267439
8.42105263157895 0.0494235449267439
8.42105263157895 0.0491198316554973
8.94736842105263 0.0491198316554973
8.94736842105263 0.0488064901137768
9.47368421052632 0.0488064901137768
9.47368421052632 0.0484876962847105
10 0.0484876962847105
10 0.0481669719627059
11 0.0481669719627059
};
\addlegendentry{Reuse}
\addplot [semithick, blue]
table {%
0 0.00996224063225974
0.526315789473684 0.00996224063225974
0.526315789473684 0.00993305406141778
1.05263157894737 0.00993305406141778
1.05263157894737 0.00992035312129491
1.57894736842105 0.00992035312129491
1.57894736842105 0.0836039721726964
2.10526315789474 0.0836039721726964
2.10526315789474 0.0839495841132843
2.63157894736842 0.0839495841132843
2.63157894736842 0.0842842409723944
3.15789473684211 0.0842842409723944
3.15789473684211 0.0844154441042306
3.68421052631579 0.0844154441042306
3.68421052631579 0.0556897277706838
4.21052631578947 0.0556897277706838
4.21052631578947 0.0554456104152286
4.73684210526316 0.0554456104152286
4.73684210526316 0.0550684816691903
5.26315789473684 0.0550684816691903
5.26315789473684 0.0545725415753029
5.78947368421053 0.0545725415753029
5.78947368421053 0.0539764762601713
6.31578947368421 0.0539764762601713
6.31578947368421 0.0533005701603597
6.84210526315789 0.0533005701603597
6.84210526315789 0.0525648171198212
7.36842105263158 0.0525648171198212
7.36842105263158 0.0517877251842132
7.89473684210526 0.0517877251842132
7.89473684210526 0.0509856330185773
8.42105263157895 0.0509856330185773
8.42105263157895 0.0501723932533482
8.94736842105263 0.0501723932533482
8.94736842105263 0.049359351063723
9.47368421052632 0.049359351063723
9.47368421052632 0.0485554665306055
10 0.0485554665306055
10 0.047767596436685
11 0.047767596436685
};
\addlegendentry{Start}
\addplot [semithick, color0]
table {%
0 0.00957981454752899
0.526315789473684 0.00957981454752899
0.526315789473684 0.0115823467870712
1.05263157894737 0.0115823467870712
1.05263157894737 0.0204494636709079
1.57894736842105 0.0204494636709079
1.57894736842105 0.0559730093844746
2.10526315789474 0.0559730093844746
2.10526315789474 0.0822158861455204
2.63157894736842 0.0822158861455204
2.63157894736842 0.0893792425269463
3.15789473684211 0.0893792425269463
3.15789473684211 0.0830374134133017
3.68421052631579 0.0830374134133017
3.68421052631579 0.0734544065794922
4.21052631578947 0.0734544065794922
4.21052631578947 0.0654594977744119
4.73684210526316 0.0654594977744119
4.73684210526316 0.0598815137687491
5.26315789473684 0.0598815137687491
5.26315789473684 0.0563002404195813
5.78947368421053 0.0563002404195813
5.78947368421053 0.0541252855534022
6.31578947368421 0.0541252855534022
6.31578947368421 0.0528479859531564
6.84210526315789 0.0528479859531564
6.84210526315789 0.0520657867323217
7.36842105263158 0.0520657867323217
7.36842105263158 0.0514670209984598
7.89473684210526 0.0514670209984598
7.89473684210526 0.050818418385237
8.42105263157895 0.050818418385237
8.42105263157895 0.0499587850811783
8.94736842105263 0.0499587850811783
8.94736842105263 0.0487942566404952
9.47368421052632 0.0487942566404952
9.47368421052632 0.0472910224679277
10 0.0472910224679277
10 0.0454640714021188
11 0.0454640714021188
};
\addlegendentry{Smooth}
\addplot [very thick, black]
table {%
0 0.01
1 0.01
1 0.0183890990440511
1.2 0.0183890990440511
1.2 0.0268179604518561
1.4 0.0268179604518561
1.4 0.0355174882966508
1.6 0.0355174882966508
1.6 0.0443816010632889
1.8 0.0443816010632889
1.8 0.0536834099310765
2 0.0536834099310765
2 0.0626555108972291
2.2 0.0626555108972291
2.2 0.0713084110276481
2.4 0.0713084110276481
2.4 0.0795286510250701
2.6 0.0795286510250701
2.6 0.0888777554465577
2.8 0.0888777554465577
2.8 0.0967470479737341
3 0.0967470479737341
3 0.1
3.2 0.1
3.2 0.0908223478234653
3.4 0.0908223478234653
3.4 0.0854014900439411
3.6 0.0854014900439411
3.6 0.0813074399360079
3.8 0.0813074399360079
3.8 0.0759351570996025
4 0.0759351570996025
4 0.0703729891292078
4.2 0.0703729891292078
4.2 0.065787352108943
4.4 0.065787352108943
4.4 0.0606915714904508
4.6 0.0606915714904508
4.6 0.0555576667684278
4.8 0.0555576667684278
4.8 0.05010310142021
5 0.05010310142021
5 0.05
6 0.05
};
\addlegendentry{True}
\end{axis}

\end{tikzpicture}}
  \caption{The model parameters for the 3-layer soil, blocky at the top and smooth at the bottom, using the Cauchy (left) and the MGS {measure} (right).}\label{fig:3layers}
\end{figure}

The synthetic profile depicts three layers in two versions.
The first version has a clear contrast between the different layers, while the second version has a smoother transition (see Figure~\ref{fig:3layers}).
We use a total of 19 intercoil spaces, uniformly distributed between \SI{1}{\meter} and \SI{19}{\meter}.
At every position we measure the HCP and PRP components of the magnetic field.
The dipole is at a frequency of \SI{1.5}{\kilo\hertz} and we put an error of \SI{5}{\percent} on the data generated with the analytical model. This error is consistent with the manufacturer-specified measurement accuracy of the instruments used in this work.

The conductivity profile in our inversion has a total of 19 layers and a half-space.
Since the half-space starts at a depth of $\SI{10}{\meter}$, the thickness of every layer is \SI[parse-numbers = false]{\frac{10}{19}}{\meter}.
We have chosen the interfaces of the true conductivity profile at \SI{2}{\meter} and \SI{4}{\meter}, and the inverted result can as a result never fully correspond with the true profile.
Hence, we avoid the usage of any prior information, which is also the case when one is inverting real data.

In both cases and for both {measure}s, the smooth result is able to detect the second layer but either shows some oscillations (Cauchy {measure}) or smears out the second layer (MGS measure).
The blocky results obtained from the start strategy clearly shows the interface depths and gives a good estimate for all the electrical conductivities.
While the latter remains true for the reuse strategy, the transition between layer 2 and 3 is still too smooth.
Further decreasing the focusing parameter causes a clearer interface, but the data misfit increases significantly (from 1.1 to 1.8).

As one can expect for the second version, the smooth profile captures the true profile very accurately.
The blocky profiles remove the gradual transition.
The position of the first interface is for all four profiles the same, while the second interface depth differs.
With the reuse strategy, the depth is put at the end of the transition zone, while the start strategy places it more at the middle.
Note that the data misfit for the start method together with the Cauchy {measure} is relatively large, see Table~\ref{tab:misfits}.
While the profile looks at first sight identical to the profile of the same strategy with the MGS {measure}, the conductivity of the third layer is a bit different.

\begin{figure}[t]
  \centering
  \subcaptionbox{}{
\begin{tikzpicture}

\definecolor{color0}{rgb}{0.75,0,0.75}

\begin{axis}[
height=.3\textwidth,
legend cell align={left},
legend style={fill opacity=0.8, draw opacity=1, text opacity=1, draw=white!80!black},
log basis y={10},
tick align=outside,
tick pos=left,
title={Cauchy},
width=.49\textwidth,
x grid style={white!69.0196078431373!black},
xlabel={d (m)},
xmin=-0.55, xmax=11.55,
xtick style={color=black},
y grid style={white!69.0196078431373!black},
ylabel={\(\displaystyle \sigma\) (Sm\(\displaystyle ^{-1}\))},
ymin=0.005, ymax=0.1,
ymode=log,
ytick style={color=black},
ytick={0.005,0.01,0.05,0.1},
yticklabels={
  \(\displaystyle 0.005\),
  \(\displaystyle 0.01\),
  \(\displaystyle 0.05\),
  \(\displaystyle 0.1\)
}
]
\addplot [semithick, red]
table {%
0 0.0146171274051066
0.526315789473684 0.0146171274051066
0.526315789473684 0.0120208262212887
1.05263157894737 0.0120208262212887
1.05263157894737 0.0120235030075037
1.57894736842105 0.0120235030075037
1.57894736842105 0.0120298449976922
2.10526315789474 0.0120298449976922
2.10526315789474 0.0643150277357357
2.63157894736842 0.0643150277357357
2.63157894736842 0.00966619794970562
3.15789473684211 0.00966619794970562
3.15789473684211 0.00966877249365178
3.68421052631579 0.00966877249365178
3.68421052631579 0.00967835138737752
4.21052631578947 0.00967835138737752
4.21052631578947 0.00969686584421248
4.73684210526316 0.00969686584421248
4.73684210526316 0.00972536056200677
5.26315789473684 0.00972536056200677
5.26315789473684 0.00976438859126409
5.78947368421053 0.00976438859126409
5.78947368421053 0.00981427404707611
6.31578947368421 0.00981427404707611
6.31578947368421 0.00987526988905716
6.84210526315789 0.00987526988905716
6.84210526315789 0.0099476619899299
7.36842105263158 0.0099476619899299
7.36842105263158 0.0100318626866843
7.89473684210526 0.0100318626866843
7.89473684210526 0.0101285364354164
8.42105263157895 0.0101285364354164
8.42105263157895 0.0102388054328962
8.94736842105263 0.0102388054328962
8.94736842105263 0.0103646025928621
9.47368421052632 0.0103646025928621
9.47368421052632 0.0105093431804746
10 0.0105093431804746
10 0.010679020910309
11 0.010679020910309
};
\addlegendentry{Reuse}
\addplot [semithick, blue]
table {%
0 0.0143817201483233
0.526315789473684 0.0143817201483233
0.526315789473684 0.014286914136348
1.05263157894737 0.014286914136348
1.05263157894737 0.00709915635339499
1.57894736842105 0.00709915635339499
1.57894736842105 0.00710133362881572
2.10526315789474 0.00710133362881572
2.10526315789474 0.0802055979939643
2.63157894736842 0.0802055979939643
2.63157894736842 0.00772519871422269
3.15789473684211 0.00772519871422269
3.15789473684211 0.00772490660268331
3.68421052631579 0.00772490660268331
3.68421052631579 0.00772769969437751
4.21052631578947 0.00772769969437751
4.21052631578947 0.00773402959767538
4.73684210526316 0.00773402959767538
4.73684210526316 0.00774363160908795
5.26315789473684 0.00774363160908795
5.26315789473684 0.00775585154511721
5.78947368421053 0.00775585154511721
5.78947368421053 0.0077689060547575
6.31578947368421 0.0077689060547575
6.31578947368421 0.0107017643142808
6.84210526315789 0.0107017643142808
6.84210526315789 0.0107230017295333
7.36842105263158 0.0107230017295333
7.36842105263158 0.0107449676418277
7.89473684210526 0.0107449676418277
7.89473684210526 0.0107661023345957
8.42105263157895 0.0107661023345957
8.42105263157895 0.0107857832320262
8.94736842105263 0.0107857832320262
8.94736842105263 0.0108035937610841
9.47368421052632 0.0108035937610841
9.47368421052632 0.0108192667308403
10 0.0108192667308403
10 0.0108326455042186
11 0.0108326455042186
};
\addlegendentry{Start}
\addplot [semithick, color0]
table {%
0 0.0146853524187106
0.526315789473684 0.0146853524187106
0.526315789473684 0.0128673898268931
1.05263157894737 0.0128673898268931
1.05263157894737 0.00732578728977743
1.57894736842105 0.00732578728977743
1.57894736842105 0.0105769458116017
2.10526315789474 0.0105769458116017
2.10526315789474 0.0786458132228384
2.63157894736842 0.0786458132228384
2.63157894736842 0.00948440235175068
3.15789473684211 0.00948440235175068
3.15789473684211 0.00691097605886722
3.68421052631579 0.00691097605886722
3.68421052631579 0.00580592380788604
4.21052631578947 0.00580592380788604
4.21052631578947 0.00544158202145564
4.73684210526316 0.00544158202145564
4.73684210526316 0.00557784645012656
5.26315789473684 0.00557784645012656
5.26315789473684 0.00615075073828575
5.78947368421053 0.00615075073828575
5.78947368421053 0.00718546588551675
6.31578947368421 0.00718546588551675
6.31578947368421 0.00873793781515479
6.84210526315789 0.00873793781515479
6.84210526315789 0.0107859188972552
7.36842105263158 0.0107859188972552
7.36842105263158 0.0130857461138681
7.89473684210526 0.0130857461138681
7.89473684210526 0.0150766268587323
8.42105263157895 0.0150766268587323
8.42105263157895 0.0160451466743809
8.94736842105263 0.0160451466743809
8.94736842105263 0.0154945524474972
9.47368421052632 0.0154945524474972
9.47368421052632 0.0134469355696712
10 0.0134469355696712
10 0.0104282486030555
11 0.0104282486030555
};
\addlegendentry{Smooth}
\addplot [very thick, black]
table {%
0 0.0129
0.2 0.0129
0.2 0.0203
0.4 0.0203
0.4 0.0121
0.6 0.0121
0.6 0.0108
0.8 0.0108
0.8 0.0104
1 0.0104
1 0.01
1.2 0.01
1.2 0.011
1.4 0.011
1.4 0.0112
1.6 0.0112
1.6 0.011
1.8 0.011
1.8 0.0132
2 0.0132
2 0.0306
2.2 0.0306
2.2 0.067
2.4 0.067
2.4 0.0651
2.6 0.0651
2.6 0.0276
2.8 0.0276
2.8 0.0203
3 0.0203
3 0.012
3.2 0.012
3.2 0.0082
3.4 0.0082
3.4 0.0074
3.6 0.0074
3.6 0.0071
3.8 0.0071
3.8 0.007
4 0.007
4 0.007
4.2 0.007
4.2 0.007
4.4 0.007
4.4 0.0068
4.6 0.0068
4.6 0.0067
4.8 0.0067
4.8 0.0068
5 0.0068
5 0.0095
5.2 0.0095
5.2 0.0089
5.4 0.0089
5.4 0.0066
5.6 0.0066
5.6 0.0096
5.8 0.0096
5.8 0.008
6 0.008
6 0.0111
6.2 0.0111
6.2 0.0145
6.4 0.0145
6.4 0.0113
6.6 0.0113
6.6 0.0101
6.8 0.0101
6.8 0.0098
7 0.0098
7 0.0096
7.2 0.0096
7.2 0.0093
7.4 0.0093
7.4 0.0094
7.6 0.0094
7.6 0.0094
7.8 0.0094
7.8 0.0096
8 0.0096
8 0.01
8.2 0.01
8.2 0.0103
8.4 0.0103
8.4 0.0109
9.4 0.0109
};
\addlegendentry{True}
\end{axis}

\end{tikzpicture}}\subcaptionbox{}{
\begin{tikzpicture}

\definecolor{color0}{rgb}{0.75,0,0.75}

\begin{axis}[
height=.3\textwidth,
legend cell align={left},
legend style={fill opacity=0.8, draw opacity=1, text opacity=1, draw=white!80!black},
log basis y={10},
tick align=outside,
tick pos=left,
title={MGS},
width=.49\textwidth,
x grid style={white!69.0196078431373!black},
xlabel={d (m)},
xmin=-0.55, xmax=11.55,
xtick style={color=black},
y grid style={white!69.0196078431373!black},
ylabel={\(\displaystyle \sigma\) (Sm\(\displaystyle ^{-1}\))},
ymin=0.005, ymax=0.1,
ymode=log,
ytick style={color=black},
ytick={0.005,0.01,0.05,0.1},
yticklabels={
  \(\displaystyle 0.005\),
  \(\displaystyle 0.01\),
  \(\displaystyle 0.05\),
  \(\displaystyle 0.1\)
}
]
\addplot [semithick, red]
table {%
0 0.014851981608212
0.526315789473684 0.014851981608212
0.526315789473684 0.0108290824948749
1.05263157894737 0.0108290824948749
1.05263157894737 0.0103048417617137
1.57894736842105 0.0103048417617137
1.57894736842105 0.0295376889631858
2.10526315789474 0.0295376889631858
2.10526315789474 0.0314407953037347
2.63157894736842 0.0314407953037347
2.63157894736842 0.0304250933089708
3.15789473684211 0.0304250933089708
3.15789473684211 0.0104175387119214
3.68421052631579 0.0104175387119214
3.68421052631579 0.00966886189882297
4.21052631578947 0.00966886189882297
4.21052631578947 0.00913147930154303
4.73684210526316 0.00913147930154303
4.73684210526316 0.00880667675204578
5.26315789473684 0.00880667675204578
5.26315789473684 0.00866477562266559
5.78947368421053 0.00866477562266559
5.78947368421053 0.00867252167307054
6.31578947368421 0.00867252167307054
6.31578947368421 0.00879964669405667
6.84210526315789 0.00879964669405667
6.84210526315789 0.00901893844791537
7.36842105263158 0.00901893844791537
7.36842105263158 0.00930496147344699
7.89473684210526 0.00930496147344699
7.89473684210526 0.00963305301991006
8.42105263157895 0.00963305301991006
8.42105263157895 0.00997900445563738
8.94736842105263 0.00997900445563738
8.94736842105263 0.0103194069227257
9.47368421052632 0.0103194069227257
9.47368421052632 0.0106324660938285
10 0.0106324660938285
10 0.0108990210928959
11 0.0108990210928959
};
\addlegendentry{Reuse}
\addplot [semithick, blue]
table {%
0 0.0148530711168121
0.526315789473684 0.0148530711168121
0.526315789473684 0.010828361209462
1.05263157894737 0.010828361209462
1.05263157894737 0.0102903194472215
1.57894736842105 0.0102903194472215
1.57894736842105 0.0295227984545457
2.10526315789474 0.0295227984545457
2.10526315789474 0.0314799177784162
2.63157894736842 0.0314799177784162
2.63157894736842 0.0305678480210591
3.15789473684211 0.0305678480210591
3.15789473684211 0.010301942002051
3.68421052631579 0.010301942002051
3.68421052631579 0.00959991964810311
4.21052631578947 0.00959991964810311
4.21052631578947 0.00909332542640478
4.73684210526316 0.00909332542640478
4.73684210526316 0.0087901633480628
5.26315789473684 0.0087901633480628
5.26315789473684 0.00866403745698189
5.78947368421053 0.00866403745698189
5.78947368421053 0.00868338455298393
6.31578947368421 0.00868338455298393
6.31578947368421 0.00881876857110267
6.84210526315789 0.00881876857110267
6.84210526315789 0.00904334876078196
7.36842105263158 0.00904334876078196
7.36842105263158 0.0093318194278388
7.89473684210526 0.0093318194278388
7.89473684210526 0.00965953275411478
8.42105263157895 0.00965953275411478
8.42105263157895 0.0100022692396461
8.94736842105263 0.0100022692396461
8.94736842105263 0.0103366553836771
9.47368421052632 0.0103366553836771
9.47368421052632 0.0106410352338186
10 0.0106410352338186
10 0.0108965246654529
11 0.0108965246654529
};
\addlegendentry{Start}
\addplot [semithick, color0]
table {%
0 0.0148824749301995
0.526315789473684 0.0148824749301995
0.526315789473684 0.0106679034665877
1.05263157894737 0.0106679034665877
1.05263157894737 0.0109526788851077
1.57894736842105 0.0109526788851077
1.57894736842105 0.0278742238082193
2.10526315789474 0.0278742238082193
2.10526315789474 0.0341508328608596
2.63157894736842 0.0341508328608596
2.63157894736842 0.0252299784487643
3.15789473684211 0.0252299784487643
3.15789473684211 0.0152546990509596
3.68421052631579 0.0152546990509596
3.68421052631579 0.0109100276355451
4.21052631578947 0.0109100276355451
4.21052631578947 0.00887066323438441
4.73684210526316 0.00887066323438441
4.73684210526316 0.00791750747532582
5.26315789473684 0.00791750747532582
5.26315789473684 0.00761282507789188
5.78947368421053 0.00761282507789188
5.78947368421053 0.00771119984560234
6.31578947368421 0.00771119984560234
6.31578947368421 0.00806401721942337
6.84210526315789 0.00806401721942337
6.84210526315789 0.00860419444729539
7.36842105263158 0.00860419444729539
7.36842105263158 0.00924505025162405
7.89473684210526 0.00924505025162405
7.89473684210526 0.00987018013674643
8.42105263157895 0.00987018013674643
8.42105263157895 0.0104387793213518
8.94736842105263 0.0104387793213518
8.94736842105263 0.0109428001116514
9.47368421052632 0.0109428001116514
9.47368421052632 0.0112071881951471
10 0.0112071881951471
10 0.0109160470254993
11 0.0109160470254993
};
\addlegendentry{Smooth}
\addplot [very thick, black]
table {%
0 0.0129
0.2 0.0129
0.2 0.0203
0.4 0.0203
0.4 0.0121
0.6 0.0121
0.6 0.0108
0.8 0.0108
0.8 0.0104
1 0.0104
1 0.01
1.2 0.01
1.2 0.011
1.4 0.011
1.4 0.0112
1.6 0.0112
1.6 0.011
1.8 0.011
1.8 0.0132
2 0.0132
2 0.0306
2.2 0.0306
2.2 0.067
2.4 0.067
2.4 0.0651
2.6 0.0651
2.6 0.0276
2.8 0.0276
2.8 0.0203
3 0.0203
3 0.012
3.2 0.012
3.2 0.0082
3.4 0.0082
3.4 0.0074
3.6 0.0074
3.6 0.0071
3.8 0.0071
3.8 0.007
4 0.007
4 0.007
4.2 0.007
4.2 0.007
4.4 0.007
4.4 0.0068
4.6 0.0068
4.6 0.0067
4.8 0.0067
4.8 0.0068
5 0.0068
5 0.0095
5.2 0.0095
5.2 0.0089
5.4 0.0089
5.4 0.0066
5.6 0.0066
5.6 0.0096
5.8 0.0096
5.8 0.008
6 0.008
6 0.0111
6.2 0.0111
6.2 0.0145
6.4 0.0145
6.4 0.0113
6.6 0.0113
6.6 0.0101
6.8 0.0101
6.8 0.0098
7 0.0098
7 0.0096
7.2 0.0096
7.2 0.0093
7.4 0.0093
7.4 0.0094
7.6 0.0094
7.6 0.0094
7.8 0.0094
7.8 0.0096
8 0.0096
8 0.01
8.2 0.01
8.2 0.0103
8.4 0.0103
8.4 0.0109
9.4 0.0109
};
\addlegendentry{True}
\end{axis}

\end{tikzpicture}}
  \caption{The model parameters with a conductivity profile based on borehole logging. Let us mention here that with the MGS {measure} the $L$-curve has the same shape using both strategies. This can also be seen in Figure~\ref{fig:embed_lcurve} where the first inversions with larger values of the focusing parameter also have the same result despite the difference in strategy.}\label{fig:luik}
\end{figure}

\begin{table}[t]
  \centering
  \caption{The data misfit for the different test cases.}\label{tab:misfits}
  \begin{tabular}{ccccccc} \toprule
    & \multicolumn{3}{c}{Cauchy} & \multicolumn{3}{c}{MGS} \\ \cmidrule(r){2-4}\cmidrule(l){5-7}
    & Smooth & Start & Reuse & Smooth & Start & Reuse \\
    \midrule
    Blocky 3 Layers & 1.06 & 1.10 & 1.09 & 1.16 & 1.17 & 1.66 \\
    Smooth 3 Layers & 1.16 & 1.43 & 1.13 & 1.10 & 1.20 & 1.16 \\
    Borehole logging & 1.08 & 1.10 & 1.12 & 1.13 & 1.13 & 1.13 \\
    Westhoek & 0.89 & 0.93 & 0.98 & 1.02 & 0.90 & 1.19 \\
    \bottomrule
  \end{tabular}
\end{table}

{While all strategies capture the global shape, there is a very clear difference if we look at the more blocky profiles. In this case the $\ell_2$-regularisation tends to oscillate, making interpretation very difficult, see for example Figure~\ref{fig:3layers}~(top) and~\ref{fig:luik}. This corroborates in general the search for an optimal focusing parameter.}

The third test is based on a conductivity profile obtained from borehole logging, for which we used the data underlying\footnote{We thank T.~Hermans for providing us with that data set.} \citep[Figure 10]{hermans_facies_2017}.
At this location, the soil consists of three layers.
From top to bottom, we find a clay layer, then a sandy gravel one and, finally, coarser gravel.
Translating this into a conductivity profile, we have a sharp peak in conductivity due to the clay layer, while below it the conductivity drops significantly due to the sandy gravel.
It then slightly increases with the coarseness of the gravel.

We use the same survey setup and noise level as in the three layers case.
The profiles obtained from our inversions are able to find the clay layer, see Figure~\ref{fig:luik}.
The MGS {measure}, however, underestimates the peak value of the conductivity.
Furthermore, the conductivity of the sandy gravel is overestimated, causing a small transition to the coarser gravel.
Note that a lower value for the focusing parameter completely removes the difference between the two lowest layers.
The reuse strategy combined with the Cauchy {measure} also fails to get the difference between the lowest regions, but has a better value for the conductivity of the clay layer.
The start strategy together with the Cauchy {measure} produces the best result, obtaining a good estimate for the peak value, whilst still distinguishing the two gravel layers.

The profiles obtained with the Cauchy {measure} are an example where the knowledge of the different strategies can be combined.
  All three solutions predict a thin conductive layer ($\sim\sigma=\SI{70}{\milli\siemens\per\meter}$, $\text{width}=\SI{0.5}{\meter}$) starting at a depth of $\SI{2}{\meter}$.
  Below this peak, there is a more resistive region, but the different strategies produce slightly different results.
  The smooth solution has some oscillations which either signals the presence of two layers with a different conductivity or are caused by the large jump in conductivity.
  The former is supported by the start strategy while the latter is indicated by the reuse strategy.
  Due to the nature of the sparse inversion, it is unlikely that a layer is introduced without a reason.

Finally, we discuss another synthetic example which mimics the EM34 instrument (Geonics), which is also used in the field data case. We use a total of 3 intercoil spacings (10 m, 20 m and 40 m), and at every position we measure the HCP and VCP components of the magnetic field, at frequencies of 6.4 kHz, 1.6 kHz and 0.4 kHz respectively. We put an error of 5\% on the data generated with the analytical model.
Moreover, we consider a more challenging problem: fitting a resistive layer of 10 m thickness (0.005 S/m) embedded in a more conductive background (0.1 S/m). Due to the larger depth of investigation associated with the EM34 setup, the discretisation reaches deeper, down to 60 m. Because of the relatively thin resistive target, we discretise the model space using 40 equidistant layers.
For simplicity, we focus here only on the Cauchy measure combined with the reuse strategy. The results for the start strategy and/or the MGS measure are omitted.

\begin{figure}
    \centering
    \includegraphics[width=0.6\linewidth]{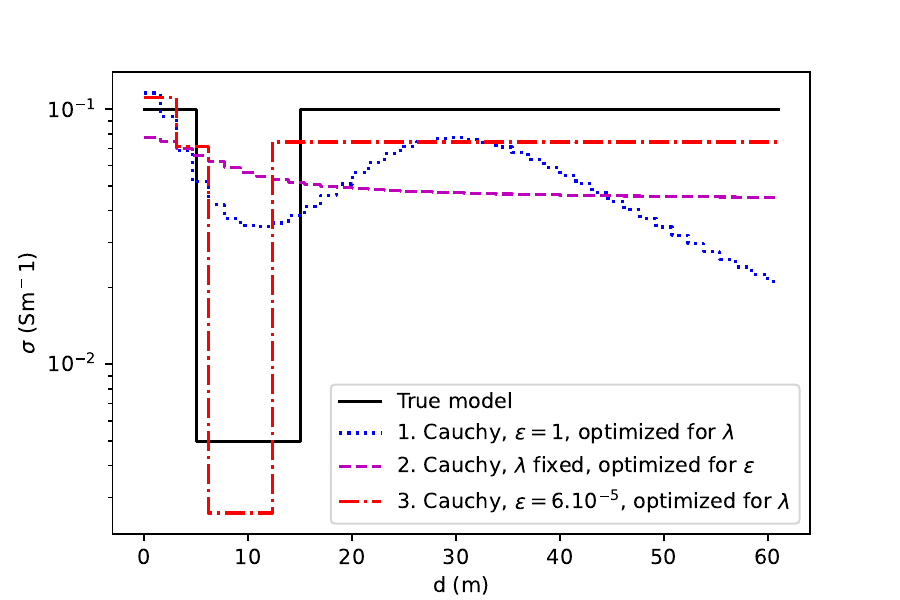}
    \caption{Inversion results for a synthetic case featuring a resistive layer (\SI{0.005}{\siemens\per\meter}, \SI{10}{\meter} thick) embedded in a conductive background (\SI{0.1}{\siemens\per\meter}) (Cauchy measure, reuse strategy).}
    \label{fig:fourthcase}
\end{figure}

The results are presented in Figure \ref{fig:fourthcase}. We first optimize for the regularisation parameter $\lambda$ and obtain the smooth model shown as the blue dotted line, corresponding to $\epsilon = 1$, with a data misfit of 0.97. We then fix this value of $\lambda$ and optimize $\epsilon$ following the procedure described above, yielding $\epsilon = 5.72 \times 10^{-5}$, shown as the magenta dashed line in Figure \ref{fig:fourthcase}. Interestingly, we observe that the data misfit has now increased to 4.07. As the final model must, of course, still fit the observed data within the discrepancy principle, this warrants a second optimization of $\lambda$.
Fixing $\epsilon$ at the value obtained above and re-optimizing $\lambda$ yields the final model (red dash-dotted line in Figure \ref{fig:fourthcase}), with a data misfit of 1.07.
This case nicely illustrates the advantage of our proposed approach: a smooth regularization can never recover the low conductivity of the resistive target, whereas our strategy successfully resolves it. We note that the final model slightly overshoots the true conductivity and recovers a target that is somewhat too narrow, an acceptable result given that only 6 data points are available in this configuration.

\subsection{Inversion of real data}

\begin{figure}[t]
  \centering
  \subcaptionbox{}{
\begin{tikzpicture}

\definecolor{darkgray176}{RGB}{176,176,176}
\definecolor{lightgray204}{RGB}{204,204,204}

\begin{axis}[
height=.3\textwidth,
legend cell align={left},
legend style={fill opacity=0.8, draw opacity=1, text opacity=1, draw=lightgray204},
log basis y={10},
tick align=outside,
tick pos=left,
width=.49\textwidth,
x grid style={darkgray176},
xlabel={d (m)},
xmin=0, xmax=40,
xtick style={color=black},
y grid style={darkgray176},
ylabel={\(\displaystyle \sigma\) (Sm\(\displaystyle ^{-1}\))},
ymin=0.001, ymax=1,
ymode=log,
ytick style={color=black}
]
\addplot [semithick, red]
table {%
0 0.003
0.2 0.003
0.2 0.004
0.4 0.004
0.4 0.005
0.600000000000001 0.005
0.600000000000001 0.007
0.8 0.007
0.8 0.007
1 0.007
1 0.007
1.2 0.007
1.2 0.008
1.4 0.008
1.4 0.012
1.6 0.012
1.6 0.012
1.8 0.012
1.8 0.013
2 0.013
2 0.013
2.2 0.013
2.2 0.014
2.4 0.014
2.4 0.02
2.6 0.02
2.6 0.027
2.8 0.027
2.8 0.029
3 0.029
3 0.026
3.2 0.026
3.2 0.025
3.4 0.025
3.4 0.02
3.6 0.02
3.6 0.018
3.8 0.018
3.8 0.018
4 0.018
4 0.018
4.2 0.018
4.2 0.019
4.4 0.019
4.4 0.019
4.6 0.019
4.6 0.02
4.8 0.02
4.8 0.022
5 0.022
5 0.024
5.2 0.024
5.2 0.029
5.4 0.029
5.4 0.035
5.6 0.035
5.6 0.048
5.8 0.048
5.8 0.063
6 0.063
6 0.081
6.2 0.081
6.2 0.097
6.4 0.097
6.4 0.112
6.6 0.112
6.6 0.124
6.8 0.124
6.8 0.136
7 0.136
7 0.15
7.2 0.15
7.2 0.165
7.4 0.165
7.4 0.18
7.6 0.18
7.6 0.193
7.8 0.193
7.8 0.209
8 0.209
8 0.225
8.2 0.225
8.2 0.244
8.4 0.244
8.4 0.255
8.6 0.255
8.6 0.264
8.8 0.264
8.8 0.275
9 0.275
9 0.283
9.2 0.283
9.2 0.289
9.4 0.289
9.4 0.298
9.6 0.298
9.6 0.308
9.8 0.308
9.8 0.318
10 0.318
10 0.325
10.2 0.325
10.2 0.335
10.4 0.335
10.4 0.343
10.6 0.343
10.6 0.35
10.8 0.35
10.8 0.354
11 0.354
11 0.352
11.2 0.352
11.2 0.339
11.4 0.339
11.4 0.327
11.6 0.327
11.6 0.308
11.8 0.308
11.8 0.285
12 0.285
12 0.253
12.2 0.253
12.2 0.225
12.4 0.225
12.4 0.2
12.6 0.2
12.6 0.183
12.8 0.183
12.8 0.165
13 0.165
13 0.152
13.2 0.152
13.2 0.132
13.4 0.132
13.4 0.103
13.6 0.103
13.6 0.079
13.8 0.079
13.8 0.061
14 0.061
14 0.047
14.2 0.047
14.2 0.038
14.4 0.038
14.4 0.034
14.6 0.034
14.6 0.031
14.8 0.031
14.8 0.028
15 0.028
15 0.026
15.2 0.026
15.2 0.025
15.4 0.025
15.4 0.024
15.6 0.024
15.6 0.023
15.8 0.023
15.8 0.022
16 0.022
16 0.021
16.2 0.021
16.2 0.021
16.4 0.021
16.4 0.021
16.6 0.021
16.6 0.021
16.8 0.021
16.8 0.022
17 0.022
17 0.021
17.2 0.021
17.2 0.02
17.4 0.02
17.4 0.018
17.6 0.018
17.6 0.018
17.8 0.018
17.8 0.018
18 0.018
18 0.017
18.2 0.017
18.2 0.016
18.4 0.016
18.4 0.015
18.6 0.015
18.6 0.014
18.8 0.014
18.8 0.014
19 0.014
19 0.014
19.2 0.014
19.2 0.014
19.4 0.014
19.4 0.014
19.6 0.014
19.6 0.014
19.8 0.014
19.8 0.014
20 0.014
20 0.014
20.2 0.014
20.2 0.014
20.4 0.014
20.4 0.015
20.6 0.015
20.6 0.015
20.8 0.015
20.8 0.015
21 0.015
21 0.015
21.2 0.015
21.2 0.015
21.4 0.015
21.4 0.016
21.6 0.016
21.6 0.016
21.8 0.016
21.8 0.016
22 0.016
22 0.016
22.2 0.016
22.2 0.016
22.4 0.016
22.4 0.016
22.6 0.016
22.6 0.016
22.8 0.016
22.8 0.016
23 0.016
23 0.016
23.2 0.016
23.2 0.016
23.4 0.016
23.4 0.016
23.6 0.016
23.6 0.016
23.8 0.016
23.8 0.016
24 0.016
24 0.016
24.2 0.016
24.2 0.016
24.4 0.016
24.4 0.016
24.6 0.016
24.6 0.016
24.8 0.016
24.8 0.016
25 0.016
25 0.016
25.2 0.016
25.2 0.015
25.4 0.015
25.4 0.015
25.6 0.015
25.6 0.015
25.8 0.015
25.8 0.015
26 0.015
26 0.015
26.2 0.015
26.2 0.015
26.4 0.015
26.4 0.015
26.6 0.015
26.6 0.015
26.8 0.015
26.8 0.015
27 0.015
27 0.015
27.2 0.015
27.2 0.015
27.4 0.015
27.4 0.015
27.6 0.015
27.6 0.015
27.8 0.015
27.8 0.014
28 0.014
28 0.014
28.2 0.014
28.2 0.014
28.4 0.014
28.4 0.014
28.6 0.014
28.6 0.013
28.8 0.013
28.8 0.013
29 0.013
29 0.014
29.2 0.014
29.2 0.014
29.4 0.014
29.4 0.014
29.6 0.014
29.6 0.015
29.8 0.015
29.8 0.015
30 0.015
30 0.015
30.2 0.015
30.2 0.015
30.4 0.015
30.4 0.015
30.6 0.015
30.6 0.015
31.6 0.015
};
\addlegendentry{borehole P11}
\addplot [semithick, blue, dashed]
table {%
0 0.003
0.2 0.003
0.2 0.003
0.4 0.003
0.4 0.003
0.600000000000001 0.003
0.600000000000001 0.003
0.800000000000001 0.003
0.800000000000001 0.004
1 0.004
1 0.004
1.2 0.004
1.2 0.005
1.4 0.005
1.4 0.006
1.6 0.006
1.6 0.007
1.8 0.007
1.8 0.008
2 0.008
2 0.009
2.2 0.009
2.2 0.009
2.4 0.009
2.4 0.01
2.6 0.01
2.6 0.011
2.8 0.011
2.8 0.011
3 0.011
3 0.012
3.2 0.012
3.2 0.012
3.4 0.012
3.4 0.012
3.6 0.012
3.6 0.012
3.8 0.012
3.8 0.012
4 0.012
4 0.012
4.2 0.012
4.2 0.012
4.4 0.012
4.4 0.012
4.6 0.012
4.6 0.013
4.8 0.013
4.8 0.013
5 0.013
5 0.013
5.2 0.013
5.2 0.013
5.4 0.013
5.4 0.013
5.6 0.013
5.6 0.013
5.8 0.013
5.8 0.013
6 0.013
6 0.013
6.2 0.013
6.2 0.012
6.4 0.012
6.4 0.012
6.6 0.012
6.6 0.012
6.8 0.012
6.8 0.012
7 0.012
7 0.013
7.2 0.013
7.2 0.013
7.4 0.013
7.4 0.014
7.6 0.014
7.6 0.015
7.8 0.015
7.8 0.016
8 0.016
8 0.016
8.2 0.016
8.2 0.018
8.4 0.018
8.4 0.019
8.6 0.019
8.6 0.021
8.8 0.021
8.8 0.023
9 0.023
9 0.026
9.2 0.026
9.2 0.031
9.4 0.031
9.4 0.041
9.6 0.041
9.6 0.054
9.8 0.054
9.8 0.079
10 0.079
10 0.116
10.2 0.116
10.2 0.149
10.4 0.149
10.4 0.193
10.6 0.193
10.6 0.221
10.8 0.221
10.8 0.246
11 0.246
11 0.268
11.2 0.268
11.2 0.286
11.4 0.286
11.4 0.303
11.6 0.303
11.6 0.316
11.8 0.316
11.8 0.32
12 0.32
12 0.311
12.2 0.311
12.2 0.287
12.4 0.287
12.4 0.249
12.6 0.249
12.6 0.199
12.8 0.199
12.8 0.157
13 0.157
13 0.118
13.2 0.118
13.2 0.087
13.4 0.087
13.4 0.068
13.6 0.068
13.6 0.053
13.8 0.053
13.8 0.045
14 0.045
14 0.038
14.2 0.038
14.2 0.032
14.4 0.032
14.4 0.027
14.6 0.027
14.6 0.025
14.8 0.025
14.8 0.024
15 0.024
15 0.023
15.2 0.023
15.2 0.023
15.4 0.023
15.4 0.023
15.6 0.023
15.6 0.022
15.8 0.022
15.8 0.022
16 0.022
16 0.022
16.2 0.022
16.2 0.022
16.4 0.022
16.4 0.022
16.6 0.022
16.6 0.021
16.8 0.021
16.8 0.021
17 0.021
17 0.021
17.2 0.021
17.2 0.02
17.4 0.02
17.4 0.02
17.6 0.02
17.6 0.02
17.8 0.02
17.8 0.02
18 0.02
18 0.019
18.2 0.019
18.2 0.018
18.4 0.018
18.4 0.017
18.6 0.017
18.6 0.016
18.8 0.016
18.8 0.015
19 0.015
19 0.014
19.2 0.014
19.2 0.013
19.4 0.013
19.4 0.013
19.6 0.013
19.6 0.013
19.8 0.013
19.8 0.013
20 0.013
20 0.013
20.2 0.013
20.2 0.013
20.4 0.013
20.4 0.013
20.6 0.013
20.6 0.014
20.8 0.014
20.8 0.014
21 0.014
21 0.014
21.2 0.014
21.2 0.014
21.4 0.014
21.4 0.013
21.6 0.013
21.6 0.013
21.8 0.013
21.8 0.013
22 0.013
22 0.014
22.2 0.014
22.2 0.014
22.4 0.014
22.4 0.013
22.6 0.013
22.6 0.013
22.8 0.013
22.8 0.013
23 0.013
23 0.014
23.2 0.014
23.2 0.014
23.4 0.014
23.4 0.014
23.6 0.014
23.6 0.014
23.8 0.014
23.8 0.014
24 0.014
24 0.014
24.2 0.014
24.2 0.014
24.4 0.014
24.4 0.013
24.6 0.013
24.6 0.013
24.8 0.013
24.8 0.013
25 0.013
25 0.013
25.2 0.013
25.2 0.013
25.4 0.013
25.4 0.013
25.6 0.013
25.6 0.013
25.8 0.013
25.8 0.013
26 0.013
26 0.013
26.2 0.013
26.2 0.013
26.4 0.013
26.4 0.013
26.6 0.013
26.6 0.013
26.8 0.013
26.8 0.013
27 0.013
27 0.013
27.2 0.013
27.2 0.013
27.4 0.013
27.4 0.013
27.6 0.013
27.6 0.013
27.8 0.013
27.8 0.013
28 0.013
28 0.013
28.2 0.013
28.2 0.013
28.4 0.013
28.4 0.013
28.6 0.013
28.6 0.013
28.8 0.013
28.8 0.013
29 0.013
29 0.013
29.2 0.013
29.2 0.013
29.4 0.013
29.4 0.013
29.6 0.013
29.6 0.013
29.8 0.013
29.8 0.013
30 0.013
30 0.013
30.2 0.013
30.2 0.013
30.4 0.013
30.4 0.013
30.6 0.013
30.6 0.013
30.8 0.013
30.8 0.014
31 0.014
31 0.014
31.2 0.014
31.2 0.015
32.2 0.015
};
\addlegendentry{borehole P14}
\end{axis}

\end{tikzpicture}}\\
  \subcaptionbox{}{\includegraphics[width=.45\textwidth]{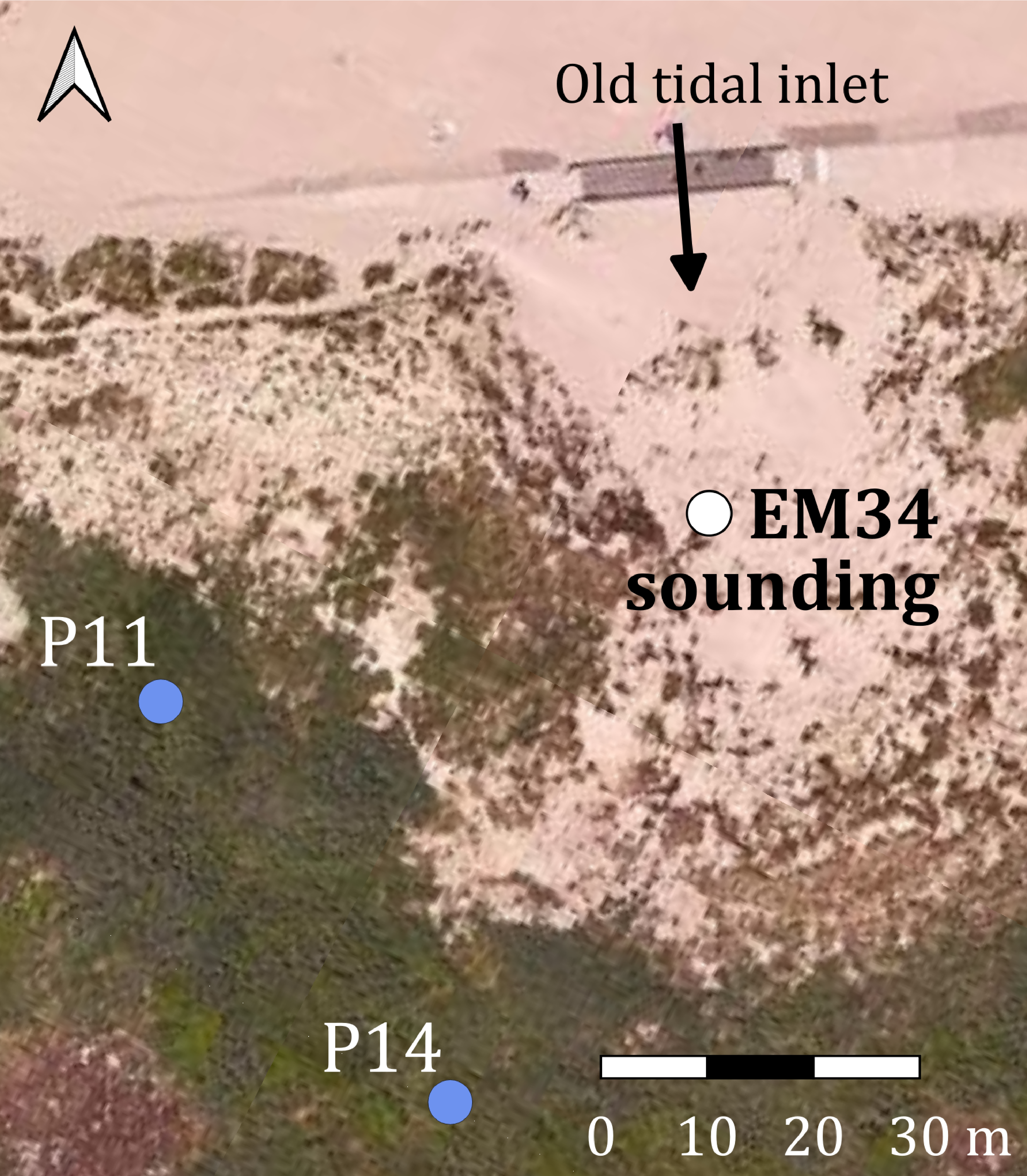}}
  \caption{Conductivity profiles obtained from borehole logging in 2012 \citep{hermans_imaging_2012}. Both wells are in the vicinity of the survey.  \label{fig:borehole} }
\end{figure}

\begin{figure}[t]
  \centering
 \includegraphics[width=\textwidth]{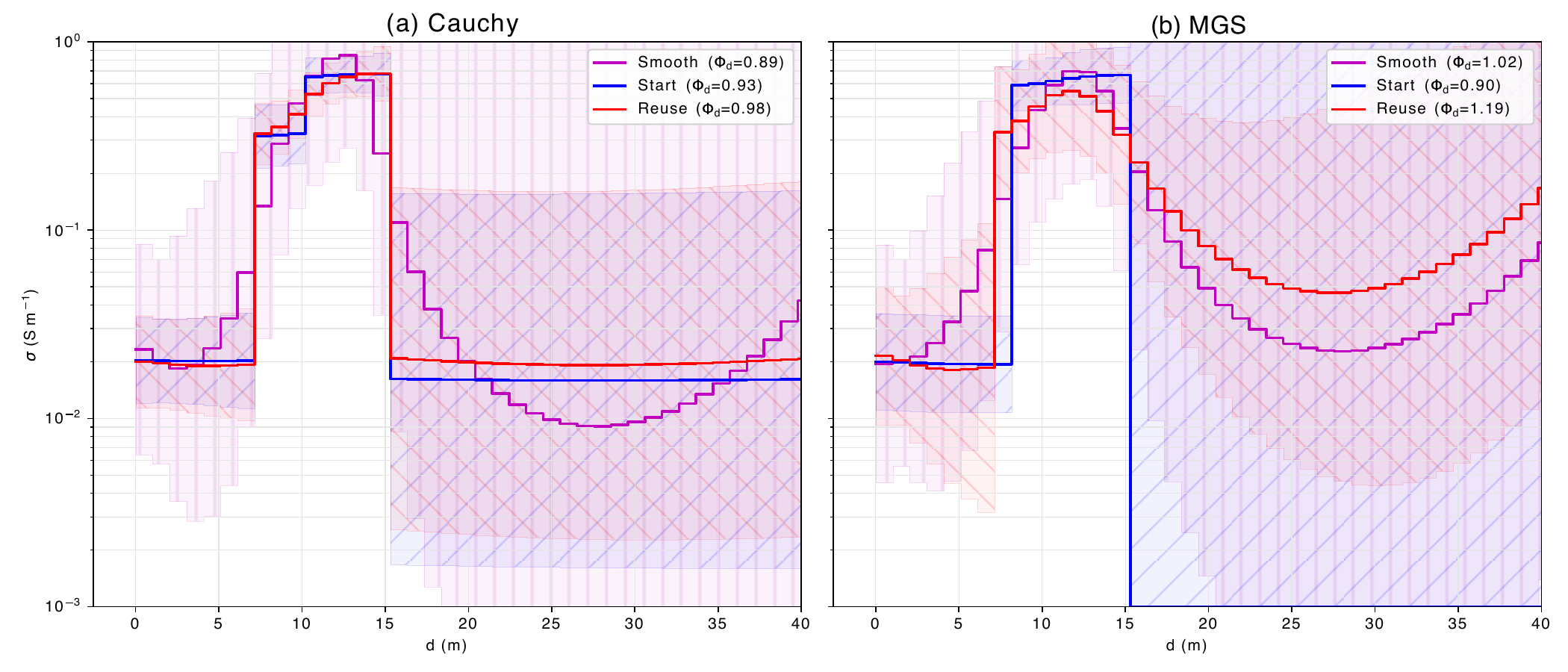}
  \caption{Inversion results for the data collected at De Westhoek nature reserve. Both measures and strategy show a sharp rise in conductivity at \SI{7}{\meter}, in accordance with prior surveys \citep{vandenbohede_salt_2008,hermans_imaging_2012}.}\label{fig:westhoek}
\end{figure}

\begin{figure}
    \centering
    \includegraphics[width=0.5\linewidth]{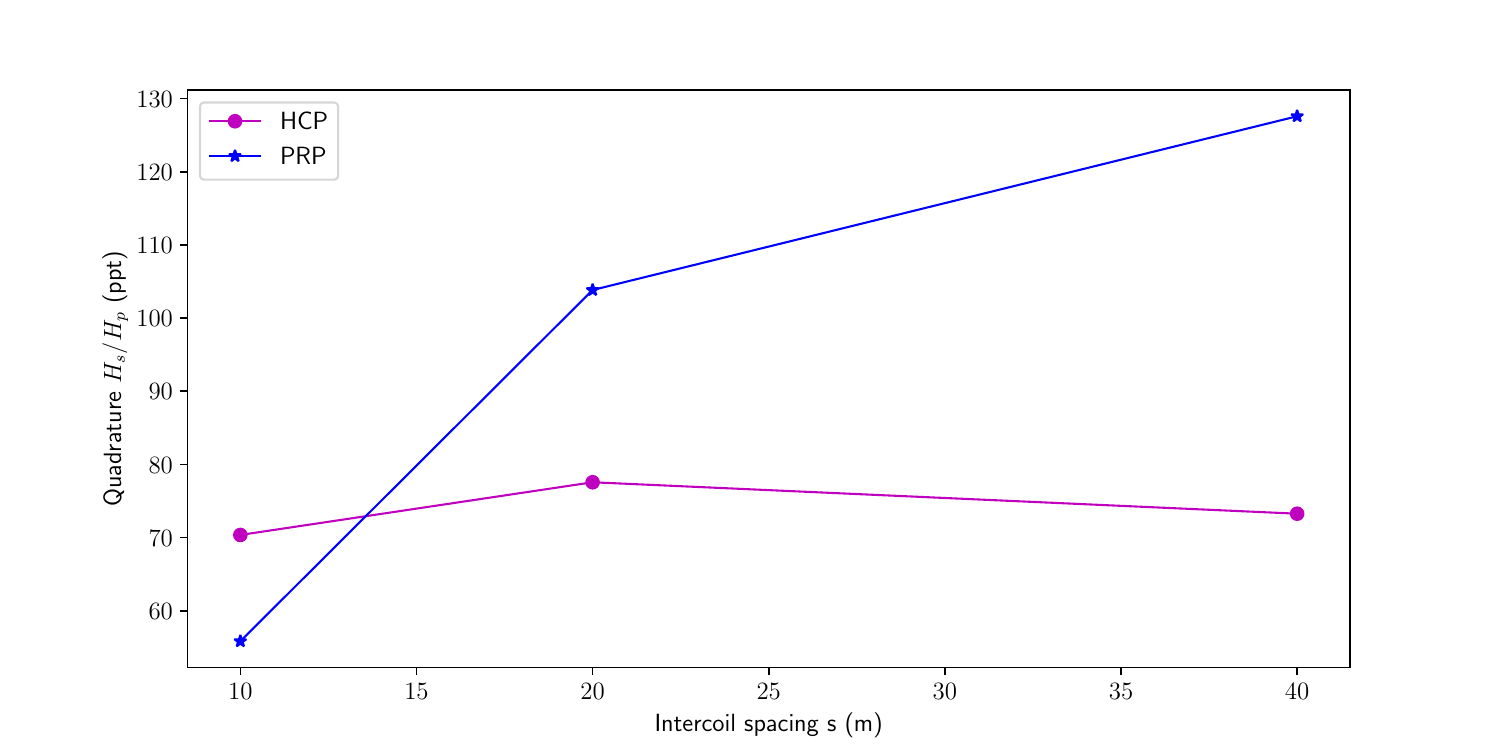}
    \caption{Observed data corresponding with Figure \ref{fig:westhoek}.}
    \label{fig:real_data_westhoek}
\end{figure}

Our final case is based on data we collected at De Westhoek nature reserve, Belgium.
This site is located at the Belgian coast and to promote bio-diversity, two inlets were created in the dunes \citep{vandenbohede_salt_2008}.
These inlets allowed seawater to enter an infiltration pond during high tide.
While these inlets have now silted up completely, there was a considerable amount of saltwater intrusion in the past.
The monitoring of this intrusion is important, because tap water is pumped up from a fresh water lens in the dune aquifer.

The soil profile consists of sand on top of the Kortrijk Formation.
The latter is a thick clay layer which starts roughly at \SI{30}{\meter} depth.
Below the water inlets, a semi-pervious layer, consisting mainly of clay, hinders the infiltration of the saltwater, causing a sharp peak in conductivity.
Based on borehole logging and electrical resistivity tomography (ERT) surveys, the highest point of the peak starts at approximately \SI{10}{\meter} depth (at well P11), it has a width of \SI{10}{\meter} and a height of \SIrange{200}{500}{\milli\siemens\per\meter} \citep{hermans_imaging_2012} (see Figure~\ref{fig:borehole}).
The height and width of the peak have changed in time due to the slow penetration of the seawater through the clay layer \citep{vandenbohede_salt_2008}.
The data was collected at the entrance of one of the inlets, we therefore expect that the intrusion is maximally visible.

Our data was collected using an EM34 apparatus \citep{geonics_geophysical_2018}.
With the EM34, measurements can be done at intercoil distances \SIlist{10;20;40}{\meter}, at a frequency of \SIlist{6.4;1.6;0.4}{\kilo\hertz} respectively.
Both the horizontal coplanar and vertical coplanar fields can be measured.
For our measurements, we assumed a data error of 5\%, the measurement error according to the instruction manual.

In Figure~\ref{fig:westhoek}, the results of our inversions are plotted.
The smooth version shows, with both {measure}s, a maximum in conductivity at \SI{12}{\meter} with a value of \SI{0.8}{\siemens\per\meter}.
After this peak, the conductivity drops to \SI{0.01}{\siemens\per\meter} and subsequently it increases gently.
The latter may indicate the presence of the Kortrijk formation, which is just within range of our survey.

Using both {measure}s, we clearly find a peak in conductivity of \SIrange{0.3}{0.6}{\siemens\per\meter} starting and ending at \SI{7}{\meter} and \SI{15}{\meter} respectively.
The characteristics of the peak are in accordance with the findings of previous surveys.
For the Cauchy {measure}, both strategies obtain almost the same result with only the conductivity of the third layer being a little bit larger than if using the reuse strategy.
The data misfit is therefore almost the same.

In case the of the MGS measure, the results of the two strategies differ significantly.
With the reuse method, a semi-smooth result is found, i.e., there is only one jump in the conductivity.
The downwards slope of the conductivity spike is a gentle decrease to a value of \SI{0.05}{\siemens\per\meter} and ends at a depth of \SI{30}{\meter}.
After reaching this low point, the conductivity increases smoothly.
Contrary to the reuse strategy, the start approach shows a clear boundary for the seawater lens.
While its width and height are in accordance with the results of other surveys, the conductivity of the third layer is clearly too low.
The data misfit of the reuse strategy is much larger than the value of the smooth result, indicating that this result is less valid.
The start method combined with the MGS measure has a similar data misfit as the Cauchy measure.

To complement the inversion results, we provide an estimate of the model uncertainty based on the linearized posterior covariance. Recall that, at convergence, the Gauss-Newton scheme already provides the approximate Hessian,
    $\mathbf{H} = \mathbf{J}^\dagger \cdot \Psi \cdot \mathbf{J}$,
which can be interpreted as the inverse of the posterior covariance matrix, $\mathbf{C}_m \approx \mathbf{H}^{-1}$. The diagonal entries of $\mathbf{C}_m$ then provide a variance estimate for each model parameter,
\begin{equation}
    \sigma(m_i) = \sqrt{(\mathbf{C}_m)_{ii}}.
    \label{eq:sigma}
\end{equation}
This estimate relies on a local Gaussian approximation of the posterior, around a linearization of the (nonlinear) forward operator. Both assumptions become less accurate for the MGS and Cauchy measures, particularly in the vicinity of the discontinuities that produce the blocky model parameters. The resulting bounds should therefore be interpreted as a lower bound on the true uncertainty, and primarily in a relative sense: they indicate which parts of the model are comparatively well- or poorly-constrained by the data, rather than providing formal confidence intervals.

Figure~\ref{fig:westhoek} shows the inversion results for De Westhoek together with these $\sigma$ bounds, for both the Cauchy and MGS measure and all three strategies (smooth, start, reuse). For the Cauchy measure, the conductivity peak associated with the saltwater intrusion, located between approximately 7 and 15~m depth, is consistently recovered by all three strategies and is tightly constrained, with narrow uncertainty bounds in this depth range. Below 15~m, the uncertainty increases substantially for the smooth and reuse results, reflecting the reduced sensitivity of the data to the deeper part of the model, while the start strategy shows an almost constant conductivity with a markedly narrower band, consistent with this strategy effectively fixing the deeper layers close to their initial value once $\epsilon$ becomes small. For the MGS measure, the same conductive peak is again well-resolved by all strategies. Markedly, the start strategy fits the data best, but has the largest uncertainty band for values at 15~m depth or larger, suggesting that this good fit is achieved in part by leaving the deeper layers relatively unconstrained.

\section{Conclusion}
\label{sec:conclusion}

We proposed two strategies to estimate the value of the focusing parameter.
In a similar way as the $L$-curve method, both strategies solve the inverse problem for different values of the focusing parameter and the optimal value is based on the plot of the model misfit in function of the data misfit.
Both strategies start with a large value of the focusing parameter, producing a smooth result.
The reuse strategy gradually lowers the value of the focusing parameter, using the previous solution as starting point for the next inverse problem.
The start strategy uses the first smooth solution as starting point for every subsequent inverse problem.
Both have a discontinuity in the curve, and we choose the result just after the discontinuity as the optimal result.

We tested our strategies with {the $\ell_0$ and  $\ell_{1}$-norm approximating functions respectively}, namely MGS and Cauchy.
From the synthetic cases, we can conclude that both strategies produce conductivity profiles sufficiently close to the true profile.
With the data we collected at De Westhoek nature reserve, a conductivity profile is constructed which is similar to profiles obtained from earlier borehole logging and ERT surveys.

With our method, it is possible to find a focusing parameter which produces a blocky profile without a priori knowledge.
  Furthermore, even if the true profile is continuous, the inversion result is still in good accordance with this.
  With the start strategy, it is also possible to obtain inversion result which are well compatible with the true profile, especially if the latter is discontinuous.
  An extra advantage of our method with respect to a priori selecting the parameter is that it prevents bias.

Since we used the damped model as forward operator in the inversions, we can conclude that the model is adequate enough to tackle the inversion.
This is further supported with the inversions where the data from our survey was used.

Future research is required to determine if our strategy is generalisable to the 2D/3D case.
Due to the possible difference in constraints between the horizontal and vertical directions, i.e.\ a smoother lateral transition versus a blocky vertical profile, it is not straightforward to extend our strategy to two different focusing parameters.

Currently, in literature, the same focusing parameter is used in both directions (out of computational simplicity we assume) for laterally constrained inversions, and we therefore expect that  we could extend our method in 2D.
  If necessary, we could rely on an already established model misfit for e.g.\ the horizontal direction (cf.\ separate regularizing term per direction), capable of giving non-blocky output in the horizontal direction.
  Our current 1D method could also first be generalized to a stitched 1D inversion.

\section*{Acknowledgments}
We are grateful to T.~Hermans for useful discussions and the providing of data sets. The research leading to these results has received funding from FWO (Fund for Scientific Research, Flanders, grant 1113020N) and from the KU Leuven Postdoctoral Mandate PDMT2/23/065.

\bibliography{paper}

@book{keller_electrical_1966,
	title = {Electrical {Methods} in {Geophysical} {Prospecting}},
	isbn = {978-0-08-011525-2},
	language = {en},
	publisher = {Elsevier Science \& Technology},
	author = {Keller, George Vernon and Frischknecht, Frank C.},
	year = {1966},
	note = {Google-Books-ID: kegNAQAAIAAJ},
}

@article{blaschek_new_2008,
  title = {A New Sensitivity-Controlled Focusing Regularization Scheme for the Inversion of Induced Polarization Data Based on the Minimum Gradient Support},
  author = {Blaschek, R. and H{\"o}rdt, A. and Kemna, A.},
  year = {2008},
  month = mar,
  journal = {Geophysics},
  volume = {73},
  number = {2},
  pages = {F45-F54},
  issn = {0016-8033, 1942-2156},
  doi = {10.1190/1.2824820},
  copyright = {\textcopyright{} 2008 Society of Exploration Geophysicists. All rights reserved.},
  langid = {english}
}

@article{deidda_frequency-domain_2022,
  title = {Frequency-{{Domain Electromagnetic Mapping}} of an {{Abandoned Waste Disposal Site}}: {{A Case}} in {{Sardinia}} ({{Italy}})},
  shorttitle = {Frequency-{{Domain Electromagnetic Mapping}} of an {{Abandoned Waste Disposal Site}}},
  author = {Deidda, Gian Piero and Himi, Mahjoub and Barone, Ilaria and Cassiani, Giorgio and Casas Ponsati, Albert},
  year = {2022},
  month = jan,
  journal = {Remote Sensing},
  volume = {14},
  number = {4},
  pages = {878},
  publisher = {{Multidisciplinary Digital Publishing Institute}},
  issn = {2072-4292},
  doi = {10.3390/rs14040878},
  copyright = {http://creativecommons.org/licenses/by/3.0/},
  langid = {english}
}

@article{deidda_inversion_2020,
  title = {Inversion of {{Multiconfiguration Complex EMI Data}} with {{Minimum Gradient Support Regularization}}: {{A Case Study}}},
  shorttitle = {Inversion of {{Multiconfiguration Complex EMI Data}} with {{Minimum Gradient Support Regularization}}},
  author = {Deidda, Gian Piero and {D{\'i}az de Alba}, Patricia and Rodriguez, Giuseppe and Vignoli, Giulio},
  year = {2020},
  month = feb,
  journal = {Mathematical Geosciences},
  issn = {1874-8961, 1874-8953},
  doi = {10.1007/s11004-020-09855-4},
  langid = {english}
}

@article{deleersnyder_inversion_2021,
  title = {Inversion of Electromagnetic Induction Data Using a Novel Wavelet-Based and Scale-Dependent Regularization Term},
  author = {Deleersnyder, Wouter and Maveau, Benjamin and Hermans, Thomas and Dudal, David},
  year = {2021},
  journal = {Geophysical Journal International},
  volume = {226},
  number = {3},
  pages = {1715--1729},
  issn = {0956-540X, 1365-246X},
  doi = {10.1093/gji/ggab182},
  copyright = {I have transferred the copyright for this publication to the publisher},
  langid = {english}
}

@article{delrue_damped_2020,
  title = {A Damped Forward {{EMI}} Model for a Horizontally Stratified Earth},
  author = {Delrue, Steven and Maveau, Benjamin and Dudal, David},
  year = {2020},
  month = jul,
  journal = {Exploration Geophysics},
  volume = {51},
  number = {4},
  pages = {422--433},
  publisher = {{Taylor \& Francis}},
  issn = {0812-3985},
  doi = {10.1080/08123985.2019.1708717}
}

@article{farquharson_constructing_2007,
  title = {Constructing Piecewise-Constant Models in Multidimensional Minimum-Structure Inversions},
  author = {Farquharson, Colin G.},
  year = {2007},
  month = dec,
  journal = {Geophysics},
  volume = {73},
  number = {1},
  pages = {K1-K9},
  issn = {0016-8033},
  doi = {10.1190/1.2816650}
}

@article{farquharson_non-linear_1998,
  title = {Non-Linear Inversion Using General Measures of Data Misfit and Model Structure},
  author = {Farquharson, Colin G. and Oldenburg, Douglas W.},
  year = {1998},
  journal = {Geophysical Journal International},
  volume = {134},
  number = {1},
  pages = {213--227},
  issn = {1365-246X},
  doi = {10.1046/j.1365-246x.1998.00555.x},
  langid = {english}
}

@article{fiandaca_generalized_2015,
  title = {Generalized Focusing of Time-Lapse Changes with Applications to Direct Current and Time-Domain Induced Polarization Inversions},
  author = {Fiandaca, Gianluca and Doetsch, Joseph and Vignoli, Giulio and Auken, Esben},
  year = {2015},
  month = nov,
  journal = {Geophysical Journal International},
  volume = {203},
  number = {2},
  pages = {1101--1112},
  issn = {0956-540X},
  doi = {10.1093/gji/ggv350},
  langid = {english}
}

@article{guitton_blocky_2012,
  title = {Blocky Regularization Schemes for {{Full-Waveform Inversion}}},
  author = {Guitton, Antoine},
  year = {2012},
  month = sep,
  journal = {Geophysical Prospecting},
  volume = {60},
  number = {5},
  pages = {870--884},
  issn = {1365-2478},
  doi = {10.1111/j.1365-2478.2012.01025.x},
  langid = {english}
}

@article{hermans_facies_2017,
  title = {Facies Discrimination with Electrical Resistivity Tomography Using a Probabilistic Methodology: Effect of Sensitivity and Regularisation},
  shorttitle = {Facies Discrimination with Electrical Resistivity Tomography Using a Probabilistic Methodology},
  author = {Hermans, Thomas and Irving, James},
  year = {2017},
  month = feb,
  journal = {Near Surface Geophysics},
  volume = {15},
  number = {1},
  pages = {13--25},
  issn = {1569-4445},
  doi = {10.3997/1873-0604.2016047},
  langid = {english}
}

@article{hermans_imaging_2012,
  title = {Imaging Artificial Salt Water Infiltration Using Electrical Resistivity Tomography Constrained by Geostatistical Data},
  author = {Hermans, Thomas and Vandenbohede, Alexander and Lebbe, Luc and Martin, Roland and Kemna, Andreas and Beaujean, Jean and Nguyen, Frederic},
  year = {2012},
  month = may,
  journal = {Journal of Hydrology},
  volume = {438--439},
  pages = {168--180},
  issn = {0022-1694},
  doi = {10.1016/j.jhydrol.2012.03.021}
}

@article{kim_time-lapse_2011,
  title = {Time-Lapse Inversion of {{2D}} Resistivity Monitoring Data with a Spatially Varying Cross-Model Constraint},
  author = {Kim, Ki-Ju and Cho, In-Ky},
  year = {2011},
  month = jul,
  journal = {Journal of Applied Geophysics},
  volume = {74},
  number = {2},
  pages = {114--122},
  issn = {0926-9851},
  doi = {10.1016/j.jappgeo.2011.04.010},
  langid = {english}
}

@article{klose_laterally_2022,
  title = {Laterally Constrained Inversion ({{LCI}}) of Multi-Configuration {{EMI}} Data with Tunable Sharpness},
  author = {Klose, Tim and Guillemoteau, Julien and Vignoli, Giulio and Tronicke, Jens},
  year = {2022},
  month = jan,
  journal = {Journal of Applied Geophysics},
  volume = {196},
  pages = {104519},
  issn = {0926-9851},
  doi = {10.1016/j.jappgeo.2021.104519},
  langid = {english}
}

@article{last_compact_1983,
  title = {Compact Gravity Inversion},
  author = {Last, B. J. and Kubik, K.},
  year = {1983},
  month = jun,
  journal = {Geophysics},
  volume = {48},
  number = {6},
  pages = {713--721},
  issn = {0016-8033},
  doi = {10.1190/1.1441501},
  langid = {english}
}

@article{loke_comparison_2003,
  title = {A Comparison of Smooth and Blocky Inversion Methods in {{2D}} Electrical Imaging Surveys},
  author = {Loke, M.H. and Acworth, Ian and Dahlin, Torleif},
  year = {2003},
  month = jun,
  journal = {Exploration Geophysics},
  volume = {34},
  number = {3},
  pages = {182--187},
  publisher = {{Taylor \& Francis}},
  issn = {0812-3985},
  doi = {10.1071/EG03182}
}

@article{martinelli_laterally_2008,
  title = {Laterally Filtered {{1D}} Inversions of Small-Loop, Frequency-Domain {{EMI}} Data from a Chemical Waste Site},
  author = {Martinelli, Patricia and Dupla{\'a}, Mar{\'i}a Celeste},
  year = {2008},
  month = jul,
  journal = {Geophysics},
  volume = {73},
  number = {4},
  pages = {F143-F149},
  publisher = {{Society of Exploration Geophysicists}},
  issn = {0016-8033},
  doi = {10.1190/1.2917197}
}

@article{more_line_1994,
  title = {Line {{Search Algorithms}} with {{Guaranteed Sufficient Decrease}}},
  author = {Mor{\'e}, Jorge J. and Thuente, David J.},
  year = {1994},
  month = sep,
  journal = {ACM Trans. Math. Softw.},
  volume = {20},
  number = {3},
  pages = {286--307},
  issn = {0098-3500},
  doi = {10.1145/192115.192132}
}

@article{morozov_solution_1966,
  title = {On the Solution of Functional Equations by the Method of Regularization},
  author = {Morozov, V.A.},
  year = {1966},
  journal = {Dokl. Akad. Nauk SSSR},
  volume = {167},
  number = {3},
  pages = {510--512},
  langid = {english}
}

@book{nocedal_numerical_2006,
  title = {Numerical Optimization},
  author = {Nocedal, Jorge and Wright, Stephen J.},
  year = {2006},
  series = {Springer Series in Operations Research},
  edition = {2nd ed},
  publisher = {{Springer}},
  address = {{New York}},
  isbn = {978-0-387-30303-1},
  langid = {english},
  lccn = {QA402.5 .N62 2006}
}

@article{paasche_cooperative_2007,
  title = {Cooperative Inversion of {{2D}} Geophysical Data Sets: {{A}} Zonal Approach Based on Fuzzy c-Means Cluster Analysis},
  shorttitle = {Cooperative Inversion of {{2D}} Geophysical Data Sets},
  author = {Paasche, Hendrik and Tronicke, Jens},
  year = {2007},
  month = mar,
  journal = {Geophysics},
  volume = {72},
  number = {3},
  pages = {A35-A39},
  issn = {0016-8033},
  doi = {10.1190/1.2670341}
}

@article{saey_electrical_2012,
  title = {Electrical {{Conductivity Depth Modelling}} with a {{Multireceiver EMI Sensor}} for {{Prospecting Archaeological Features}}},
  author = {Saey, Timothy and De Smedt, Philippe and Meerschman, Eef and Islam, Mohammad Monirul and Meeuws, Fun and Van De Vijver, Ellen and Lehouck, Alexander and Van Meirvenne, Marc},
  year = {2012},
  month = jan,
  journal = {Archaeological Prospection},
  volume = {19},
  number = {1},
  pages = {21--30},
  issn = {1099-0763},
  doi = {10.1002/arp.425},
  langid = {english}
}

@article{scudiero_constrained_2011,
  title = {Constrained Optimization of Spatial Sampling in Salt Contaminated Coastal Farmland Using {{EMI}} and Continuous Simulated Annealing.},
  author = {Scudiero, E. and Deiana, R. and Teatini, P. and Cassiani, G. and Morari, F.},
  year = {2011},
  month = jan,
  journal = {Procedia Environmental Sciences},
  series = {Spatial {{Statistics}} 2011: {{Mapping Global Change}}},
  volume = {7},
  pages = {234--239},
  issn = {1878-0296},
  doi = {10.1016/j.proenv.2011.07.041},
  langid = {english}
}

@article{thibaut_new_2021,
  title = {A New Workflow to Incorporate Prior Information in Minimum Gradient Support ({{MGS}}) Inversion of Electrical Resistivity and Induced Polarization Data},
  author = {Thibaut, Robin and Kremer, Thomas and Royen, Annie and Kim Ngun, Bun and Nguyen, Fr{\'e}d{\'e}ric and Hermans, Thomas},
  year = {2021},
  month = apr,
  journal = {Journal of Applied Geophysics},
  volume = {187},
  pages = {104286},
  issn = {0926-9851},
  doi = {10.1016/j.jappgeo.2021.104286},
  langid = {english}
}

@article{tikhonov_stability_1943,
  title = {On the Stability of Inverse Problems},
  author = {Tikhonov, A.},
  year = {1943},
  journal = {Dokl. Akad. Nauk SSSR},
  volume = {39},
  number = {5},
  pages = {195--198},
  langid = {english}
}

@article{vandenbohede_salt_2008,
  title = {Salt Water Infiltration in Two Artificial Sea Inlets in the {{Belgian}} Dune Area},
  author = {Vandenbohede, A. and Lebbe, L. and Gysens, Stefaan and Delecluyse, Kevin and DeWolf, Peter},
  year = {2008},
  month = oct,
  journal = {Journal of Hydrology},
  volume = {360},
  number = {1},
  pages = {77--86},
  issn = {0022-1694},
  doi = {10.1016/j.jhydrol.2008.07.018},
  langid = {english}
}

@book{vogel_computational_2002,
  title = {Computational {{Methods}} for {{Inverse Problems}}},
  author = {Vogel, Curtis R.},
  year = {2002},
  month = jan,
  publisher = {{SIAM}},
  googlebooks = {EhnLWkc0GwAC},
  isbn = {978-0-89871-550-7},
  langid = {english}
}

@article{wolfe_convergence_1969,
  title = {Convergence {{Conditions}} for {{Ascent Methods}}},
  author = {Wolfe, Philip},
  year = {1969},
  month = apr,
  journal = {SIAM Review},
  volume = {11},
  number = {2},
  pages = {226--235},
  publisher = {{Society for Industrial and Applied Mathematics}},
  issn = {0036-1445},
  doi = {10.1137/1011036}
}

@online{geonics_geophysical_2018,
  title = {Geophysical {{Instrumentation}} for {{Geology}}, {{Military}}, {{Environment}}, {{Agriculture}}, {{Archaeology}}, and {{Geotechnical Studies}}},
  author = {{Geonics}},
  year = {2018},
  url = {http://www.geonics.com/html/products.html},
  urldate = {2018-02-04}
}

@article{wait_note_1962,
    author = {Wait, James R.},
    title = "{A note on the electromagnetic response of a stratified Earth}",
    journal = {Geophysics},
    volume = {27},
    number = {3},
    pages = {382-385},
    year = {1962},
    month = {06},
    issn = {0016-8033},
    doi = {10.1190/1.1439028},
    url = {https://doi.org/10.1190/1.1439028},
    eprint = {https://pubs.geoscienceworld.org/geophysics/article-pdf/27/3/382/3152958/382.pdf},
}

@inproceedings{deleersnyder2023estimating,
	title={Estimating the Volume of Recoverable Materials from a Former Gypsum Landfill Using Frequency-Domain Electromagnetics},
	author={Deleersnyder, Wouter and Bah, Abdoulie and Blanchy, Guillaume and Hermans, Thomas and Van De Vijver, Ellen},
	booktitle={NSG2023 29th European Meeting of Environmental and Engineering Geophysics},
	volume={2023},
	number={1},
	pages={1--5},
	year={2023},
	organization={European Association of Geoscientists \& Engineers}
}

@inproceedings{deidda2018smooth,
	title={Smooth and sparse inversion of EMI data from multi-configuration measurements},
	author={Deidda, Gian Piero and de Alba, Patricia Diaz and Vignoli, Giulio and Rodriguez, Giuseppe},
	booktitle={2018 IEEE 4th International Forum on Research and Technology for Society and Industry (RTSI)},
	pages={1--6},
	year={2018},
	organization={IEEE}
}

@article{zhdanov2004minimum,
	title={Minimum support nonlinear parametrization in the solution of a 3D magnetotelluric inverse problem},
	author={Zhdanov, Michael and Tolstaya, Ekaterina},
	journal={Inverse problems},
	volume={20},
	number={3},
	pages={937},
	year={2004},
	publisher={IOP Publishing}
}

@article{carbajal2012focused,
	title={Focused time-lapse inversion of radio and audio magnetotelluric data},
	author={Carbajal, Marina Rosas and Linde, Niklas and Kalscheuer, Thomas},
	journal={Journal of Applied Geophysics},
	volume={84},
	pages={29--38},
	year={2012},
	publisher={Elsevier}
}

@article{hansen1999curve,
	title={The L-curve and its use in the numerical treatment of inverse problems},
	author={Hansen, Per Christian},
	year={1999},
	publisher={IMM, Department of Mathematical Modelling, Technical Universityof Denmark~…}
}

@article{deleersnyder2023flexible,
  title={Flexible quasi-2D inversion of time-domain AEM data, using a wavelet-based complexity measure},
  author={Deleersnyder, Wouter and Maveau, Benjamin and Hermans, Thomas and Dudal, David},
  journal={Geophysical Journal International},
  volume={233},
  number={3},
  pages={1847--1862},
  year={2023},
  publisher={Oxford University Press}
}

@article{rulff2026computational,
  title={Computational electromagnetic geophysics for groundwater system studies: A review of established practices and recent advances},
  author={Rulff, Paula and Castillo-Reyes, Octavio and Deleersnyder, Wouter and Mascarell, Maria Carrizo and Minsley, Burke J and King, Jude},
  journal={Journal of Hydrology},
  volume={677},
  pages={135542},
  year={2026},
  publisher={Elsevier}
}

@article{feng2020modified,
  title={A modified total variation regularization approach based on the Gauss-Newton algorithm and split Bregman iteration for magnetotelluric inversion},
  author={Feng, Deshan and Su, Xuan and Wang, Xun and Wang, Xiangyu},
  journal={Journal of Applied Geophysics},
  volume={178},
  pages={104073},
  year={2020},
  publisher={Elsevier}
}
\end{document}